\documentclass[%
 reprint,
superscriptaddress,
 amsmath,amssymb,
 aps,
pra,
]{revtex4-1}

\usepackage{lipsum}
\usepackage{xcolor}
\usepackage{graphicx}
\usepackage{dcolumn}
\usepackage{bm}
\usepackage{siunitx}
\usepackage{xcolor}
\usepackage{dsfont}
\usepackage{tikz}
\usepackage{tikz-3dplot}
\usepackage{qcircuit}
\usepackage{hyperref}
\usepackage{physics}
\usepackage{bbold}
\usepackage{float}
\hypersetup{colorlinks=true,citecolor=blue,linkcolor=blue,filecolor=blue,urlcolor=blue,breaklinks=true}

\tikzset{text/.default=} 

\begin{document}

\preprint{APS/123-QED}

\title{Improving readout in quantum simulations with repetition codes}

\author{Jakob M. G\"unther}
 \affiliation{
IBM Quantum, IBM Research – Zurich, 8803 R\"uschlikon, Switzerland
}

\author{Francesco Tacchino}
 \affiliation{
IBM Quantum, IBM Research – Zurich, 8803 R\"uschlikon, Switzerland
 }
 \author{James R. Wootton}
 \affiliation{
IBM Quantum, IBM Research – Zurich, 8803 R\"uschlikon, Switzerland
 }
 \author{Ivano Tavernelli}
 \affiliation{
IBM Quantum, IBM Research – Zurich, 8803 R\"uschlikon, Switzerland
 }
 \author{Panagiotis Kl. Barkoutsos}
 \affiliation{
IBM Quantum, IBM Research – Zurich, 8803 R\"uschlikon, Switzerland
 }
 
\date{\today}

\begin{abstract}

Near term quantum computers suffer from the presence of different noise sources. 
In order to mitigate for this effect and acquire results with significantly better accuracy, there is the urge of designing efficient error correction or error mitigation schemes. 
The cost of such techniques is usually high in terms of resource requirements, either in hardware or at the algorithmic level. 
In this work, we follow a pragmatic approach and we use repetition codes as scalable schemes with the potential to provide more accurate solutions to problems of interest in quantum chemistry and physics. 
We investigate different repetition code layouts and we propose a circular repetition scheme with connectivity requirements that are native on IBM Quantum hardware. 
We showcase our approach in multiple IBM Quantum devices and validate our results using a simplified theoretical noise model. 
We highlight the effect of using the proposed scheme in an electronic structure VQE calculation and in the simulation of time evolution for a quantum Ising model.

\end{abstract}

\maketitle

\section{Introduction}

Quantum computers are composed of fragile quantum systems that cannot be fully isolated from external noise, and whose manipulations imply a level of uncertainty (unless error correction protocols are applied).
These conditions lead to errors, whenever a quantum circuit is executed on near term (i.e., non fault tolerant) devices. 
Although the exact nature of these inaccuracies depends on the different qubit technologies employed,
the presence of errors is inherent to all qubit architectures~\cite{roffe:19}.
These limitations of state-of-the-art quantum computers stands at odds with the need for accuracy in the calculations (e.g., chemical accuracy in electronic structure calculations). 
For this reason, schemes to make the computation resilient to errors are required. 

Multiple error mitigation techniques~\cite{endo2020hybrid} have been proposed in the literature. 
Such techniques can extend the range of operations for quantum computers using protocols inspired from capabilities and limitations of current hardware.
Recent schemes propose the use of zero noise extrapolation~\cite{extrapolate1,extrapolate2,extrapolate3,extrapolate4, Endo2017, Endo2019mitigating, Barron2020, Sun2021mitigating}, quasi-probability distribution~\cite{extrapolate3, Endo2017,piveteau2021error,lostaglio2021error}, symmetry enforcement~\cite{qse1, qse2, McClean2016, Sagastizabal2019, Barron2020}, stabilizer-like methods~\cite{McArdle2019, andersen2020repeated}, virtual distillation~\cite{huggins2020virtual,koczor2020exponential}
or general algorithmic schemes~\cite{Chiesa2019,suchsland2020algorithmic}. 
For the particular case of mitigating measurement errors there are certain schemes proposed for the increased accuracy in the estimation of mean values of observables~\cite{Chow2010,Ryan2015,Chen2019}.
Although these schemes extend the computational reach of current quantum processors, they just constitute an intermediate step towards fully error corrected (fault tolerant) quantum computation.
Recently, some proposals for integrating error correction and error mitigation techniques, thus moving towards fault tolerant quantum architectures, have also been put forward~\cite{Suzuki2020quantum,piveteau2021error,lostaglio2021error}.

Quantum Error Correction (QEC) codes provide a full and scalable solution for the protection of quantum information against quantum noise in a wide range of forms.
This comes at the expense of significant overheads, in terms of how many noisy qubits are required to construct one without errors, the number of noisy gates required to perform even a single gate accurately, and the native gate set demanded of the underlying architecture.
The last decades have seen a great deal of progress in finding error-correcting codes that minimize these overheads, and allow us to move towards fault-tolerant quantum computing architectures~\cite{lidar:13,roffe:19}.
Even so, the use of quantum error correction in full, quantum algorithms is beyond the capabilities of current and near-term quantum hardware.
Instead, the current focus is on small, proof-of-principle implementations, used to benchmark progress towards fault-tolerance \cite{kelly:14, riste:15, corcoles:15, linke:17, takita:17, vuillot:18, wootton:18, naveh:18, andersen:20, wootton:20, gong:21,chen:21}.

The simplest example of a classical error correcting code is the repetition code~\cite{nielsen_chuang}. Put simply, this method consists in the preparation of many copies of a given stored information. When extracting an output, a comparison of the different copies can be used to detect and discard the spurious effects of errors and deduce the original information. When used to store bit values, the value to be stored is repeated across a string of multiple bits. Errors in this case manifest as simple bit flips. The detection of errors uses the fact that all bits used in the encoding should take the same value. By comparing pairs of bits and finding instances where their values are not equal, the effects of bit flip errors can then be found.

Quantum error correction codes operate on the same principle:  qubit states are encoded in collective states of many (noisy) qubits that posses certain symmetries. 
By checking those symmetries, by means of so-called syndrome measurements, errors can be detected and corrected. 
Note that in the quantum case, errors manifest in more complex forms than simple bit flips described here. 
The repetition code on its own is therefore not sufficient to fully protect against quantum errors, and more sophisticated encoding methods are required in the most general case. 
Nevertheless, quantum implementations of the repetition code can often serve as a good starting point for quantum error correction.
Recent studies \cite{kelly:14, wootton:18, naveh:18, wootton:20, chen:21} have shown that 
repetition codes have the potential to be used with current hardware.
To date, the strongest evidences of this fact  are those by Wootton~\cite{wootton:20} and Chen et al~\cite{chen:21}. 
The former is the largest in terms of number of qubits, implementing a repetition code in a linear chain of 43 qubits. The latter is the largest in terms of 
syndrome measurement rounds, showing that error suppression is sustained over 50 rounds. 

The possibility of combining those strategies with applications of near term quantum computers in order to yield results with better accuracy remains an open research question that we try to address in this work. 
The paper is organised as follows. 
In Sec.~\ref{sec:theory} we discuss different repetition encodings and we provide insights of the theoretical noise model that we used to assess the performances of our schemes.
In Sec.~\ref{sec:results} we test our proposed protocols on the state-of-the-art superconducting quantum hardware provided by IBM Quantum and we demonstrate the effectiveness of repetition codes in mitigating some classes of logical errors. 
We also compare to the theoretical noise model calculations and provide upper and lower bounds for the logical errors.
We apply our technique to the solution of quantum chemistry and physics models and highlight the advantageous properties of application of repetition codes. 
We conclude in Sec.~\ref{sec:conclusions} with some important remarks related to the near-term applicability and the scalability of our approach for future quantum simulations. 

\section{Theory}
\label{sec:theory}

\subsection{Repetition codes} 
\label{sec:repetition_codes}

Similarly to their classical counterparts, quantum repetition codes leverage redundancy to protect information from the effects of noise. In this work, we introduce an application of those techniques to the mitigation of readout errors in current superconducting quantum hardware. 
Since one of the most dominant sources of inaccuracies affecting 
readout operations is most often represented by bit flips in the computational basis 
in the following we will focus on repetition encoding schemes specifically designed to detect and correct this form of errors. 

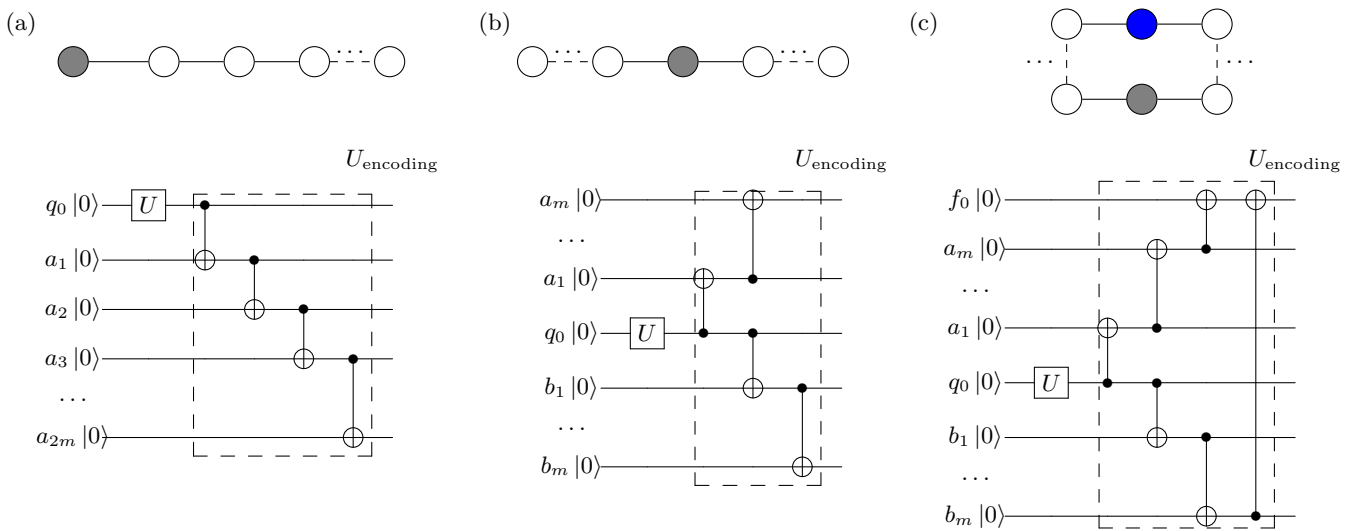
\begin{figure*}[htb!]
    \centering
    \begin{tikzpicture}
\node[text] at (-0.9,0.5) {(a)};
\node[anchor = east, circle,draw, minimum size=0.4cm, fill = gray] (A1) at  (0,0) {};
\node[circle,draw, minimum size=0.4cm] (B1) at  (1,0)  {};
\node[circle,draw, minimum size=0.4cm] (C1) at  (2,0)  {};
\node[circle,draw, minimum size=0.4cm] (D1) at  (3,0)  {};
\node[circle,draw, minimum size=0.4cm] (E1) at  (4,0)  {};
\draw (A1) -- (B1);
\draw (B1) -- (C1);
\draw (C1) -- (D1);
\draw [dashed] (D1) --node[above]{\dots} (E1);

\node[anchor=north] (A1) at  (2.5,-1.2) {
\Qcircuit @C=1.2em @R=1.2em {
& & & & & & & & \mbox{$U_{\rm{encoding}}$} & & & &\\
&q_0\ket{0}  & \quad & \gate{U} & \ctrl{1} & \qw & \qw & \qw & \qw \\
&a_1\ket{0}  & \quad & \qw & \targ & \ctrl{1}& \qw & \qw & \qw  \\
&a_2\ket{0}  & \quad & \qw & \qw & \targ & \ctrl{1} & \qw & \qw \\
&a_3\ket{0}  & \quad & \qw & \qw & \qw & \targ & \ctrl{2} & \qw\\\
& \dots \\
&a_{2m}\ket{0}  & \quad & \qw & \qw & \qw & \qw &\targ & \qw \gategroup{2}{5}{7}{8}{.7em}{--}
}
};

\node[text] at (5.4,0.5) {(b)};
\node[circle,draw, minimum size=0.4cm] (A2) at  (5.9,0) {};
\node[circle,draw, minimum size=0.4cm] (B2) at  (6.9,0)  {};
\node[circle,draw, minimum size=0.4cm, fill = gray] (C2) at  (7.9,0)  {};
\node[circle,draw, minimum size=0.4cm] (D2) at  (8.9,0)  {};
\node[circle,draw, minimum size=0.4cm] (E2) at  (9.9,0)  {};
\draw [dashed] (A2) --node[above]{\dots} (B2);
\draw (B2) -- (C2);
\draw (C2) -- (D2);
\draw [dashed] (D2) --node[above]{\dots} (E2);

\node[anchor=north] (A1) at  (8.8,-1.2) {
\Qcircuit @C=1.2em @R=1.2em {
& & & & & & & \mbox{$U_{\rm{encoding}}$} & & & &\\
&a_m\ket{0}  & \quad & \qw & \qw & \targ& \qw & \qw   \\
&\dots \\
&a_1\ket{0}  & \quad & \qw & \targ & \ctrl{-2} & \qw & \qw  \\
&q_0\ket{0}  & \quad & \gate{U} & \ctrl{-1} & \ctrl{1} & \qw & \qw \\
&b_1\ket{0}  & \quad & \qw & \qw  & \targ & \ctrl{2} & \qw\\
&\dots \\
&b_m\ket{0}  & \quad & \qw & \qw & \qw &\targ & \qw \gategroup{2}{5}{8}{7}{.7em}{--}
}
};

\node[text] at (11.1,0.5) {(c)};
\node[circle,draw, minimum size=0.4cm] (A3) at  (13,-0.5) {};
\node[circle,draw, minimum size=0.4cm, fill = gray] (B3) at  (14,-0.5)  {};
\node[circle,draw, minimum size=0.4cm] (C3) at  (15,-0.5)  {};
\node[circle,draw, minimum size=0.4cm] (D3) at  (13,0.5)  {};
\node[circle,draw, minimum size=0.4cm, fill = blue] (E3) at  (14,0.5)  {};
\node[circle,draw, minimum size=0.4cm] (F3) at  (15,0.5)  {};
\draw (A3) -- (B3);
\draw [dashed] (A3) --node[left]{\dots} (D3);
\draw (B3) -- (C3);
\draw (D3) -- (E3);
\draw (E3) -- (F3);
\draw [dashed] (C3) --node[right]{\dots} (F3);

\node[anchor=north] (A1) at  (14.5,-1.2) {
\Qcircuit @C=1.2em @R=1.2em {
& & & & & & & & \mbox{$U_{\rm{encoding}}$} & & & &\\
&f_0\ket{0}  & \quad & \qw & \qw & \qw & \targ & \targ &\qw  \\
&a_{m}\ket{0}  & \quad & \qw & \qw & \targ& \ctrl{-1} &\qw & \qw \\
&\dots \\
&a_1\ket{0}  & \quad & \qw & \targ & \ctrl{-2} & \qw & \qw & \qw \\
&q_0\ket{0}  & \quad & \gate{U} & \ctrl{-1} & \ctrl{1} &  \qw & \qw & \qw \\
&b_1\ket{0}  & \quad & \qw & \qw & \targ & \ctrl{2} & \qw & \qw\\
&\dots \\
&b_{m}\ket{0}  & \quad & \qw & \qw & \qw &\targ & \ctrl{-7} &\qw \gategroup{2}{5}{9}{8}{.7em}{--}
}
};

\end{tikzpicture}
    \caption{Repetition code layouts with $2m = n_{rep}$ auxiliary qubits.
    (a) \textit{Chain Encoding.} 
    The physical qubit (grey) is connected to auxiliary qubits (white) with a ladder of CNOTs in a chain. (b) \textit{Split Encoding.} The physical qubit (grey) is connected with two branches of auxiliary qubits (white) with two independent ladders of CNOTs. (c) \textit{Circular Encoding.} The two branches of auxiliary qubits (white) starting from the physical qubit(grey) are connect to the flag qubit (blue) closing the loop of qubits. The two ladder of CNOTs are followed by 2 CNOT gates connecting the flag qubit with the last qubit of the two branches.}
    \label{fig:Circuit_encoding_diagram}
\end{figure*}

Let us assume that a single physical qubit, hereafter named the \textit{root qubit}, is given in a generic state $a\ket{0}+b\ket{1}$. By using $n_{rep}$ additional physical qubits, initially prepared in state $\ket{0}$, we perform the mapping 

\begin{equation}
    (a\ket{0}+b\ket{1})\ket{0\hdots 0} \mapsto a\ket{0\hdots 0} + b\ket{1\hdots 1}
\end{equation}
Such an operation can be implemented in practice by applying, for example, $n_{rep}$ CNOT gates targeting the $n_{rep}$ auxiliary qubits as in Figure~\ref{fig:Circuit_encoding_diagram}.

From a formal perspective, this encoding is generated by the $n_{rep}$ stabilizers $\{\mathrm{Z}_0\mathrm{Z}_1, \mathrm{Z}_1\mathrm{Z}_2,...,\mathrm{Z}_{n_{rep}-1}\mathrm{Z}_{n_{rep}}\}$, where $\mathrm{Z}_i$ is the Pauli $\sigma_z$ quantum gate applied to qubit $i$ and we associate the index $i = 0$ to the root qubit. 
Notice that the state $a\ket{0\hdots 0} + b\ket{1\hdots 1}$ is not equivalent to $n_{rep}+1$ copies of the state $(a\ket{0}+b\ket{1})$, as this would in general violate the no-cloning theorem, and only the eigenvalues associated to the $\mathrm{Z}$ operator of the original single-qubit state are stored redundantly. 
As a result, only errors caused by the bit flip channel, associated to the Pauli $\mathrm{X}$ operator, can be detected and corrected with this encoding. 
With some modifications, it is possible to correct for analogous errors caused by the channels generated by the $\mathrm{Y}$ or $\mathrm{Z}$ operators, although we will not explore such possibilities in this work. As we concentrate on readout error mitigation, we will always assume that the encoding is performed right before the quantum measurements, which will therefore be the only operations acting on encoded states. All other operations describing the preparation of the joint state of one or more root qubits will be implemented without encoding, so that the question of how to realize fault-tolerantly the required gates in the repetition code is not touched upon here.

We recover error-mitigated readout results for individual root qubits via a majority vote procedure applied as a post-processing stage on the outcome bit strings. In particular, any bit string $b_0\ldots b_{n_{rep}}$ resulting from a quantum measurement in the computational basis is decoded as $0$ or $1$ depending on which of the two values appears more often in the $n_{rep} + 1$ output bits. The decoding is successful whenever less than $(n_{rep}+1)/2$ bit values were flipped by noise. For simplicity, in the following we will only consider repetition codes for which $n_{rep} + 1$ is odd.

The overall procedure is summarized in Fig.~\ref{fig:Circuit_encoding_diagram} for the 
single root qubit case for three different repetition layouts. 
Each root qubit is mapped to its encoded version
which involves an ancillary register and an encoding-dependent structure of controlled operations, $U_{\rm{encoding}}$. 
Finally, all qubits are measured in the computational basis and a majority vote is performed in order to recover the measurement values for the root qubit. 

\paragraph*{Repetition Code Layouts} When implemented on real noisy quantum processors, the performances of the proposed repetition-based error mitigation strategy is influenced by many different variables, such as the number of encoding qubits, their quality and connectivity on hardware and the layout of the required CNOT chains.
As a matter of fact, the practical realization of the $U_{\rm{encoding}}$ operations introduces some complexity overhead that must be weighed against the benefits of the repetition encoding itself. 

As mentioned above, the typical structure of $U_{\rm{encoding}}$ corresponds to a sequence of CNOT operations targeting auxiliary qubits in a sequential way. 
In a simple \textit{chain encoding} scenario, one uses a linear sequence of CNOTs in which all the auxiliary qubits are arranged consecutively, acting first as a target and then as a control (see 
Fig.~\ref{fig:Circuit_encoding_diagram}a). This strategy has only modest requirements in terms of hardware connectivity but may be affected by error-propagation effects lowering its performances. 
In fact, a single bit flip error occurring, e.g., on the first CNOT between the root qubit and $a_1$ will lead to a wrong outcome after decoding. 
An immediate improvement can be obtained by adopting a \textit{split encoding} approach, in which a single chain is split into branches. 
This does not only reduce the sensitivity to error-propagation effects, but allows many of the CNOT operations to be executed in parallel, therefore reducing the effective circuit depth. 
If we assume again a linear connectivity between identical qubits, there are in principle $(n_{rep})/2+1$ nonequivalent ways to realize either a \textit{chain} or \textit{split} encoding, depending on the position of the root-qubit within the chain. 
In this study we will focus on the two edge cases, with the root qubit positioned either at one end of the chain or in the middle. 

\paragraph*{Circular Repetition Code} To further improve the performances of repetition encodings, we propose a \textit{circular} variant, directly inspired by the heavy hexagonal topology of IBM Quantum processors and explicitly designed to protect against bit flip errors induced by the implementation of $U_{\rm{encoding}}$ itself. The circular repetition code derives from the split-repetition layout and utilizes one additional flag qubit, which is initialized in state $\ket{0}$. The two split-repetition branches each connect to the flag-qubit with a CNOT, where the flag qubit is the target in both cases, see Fig.~\ref{fig:Circuit_encoding_diagram}b.

As in the general case, we can write the initial unencoded state as $(a\ket{0} + b\ket{1})\ket{0\hdots 0}\ket{0\hdots 0}\ket{0}$, where the first and last qubit correspond to root and flag qubit respectively, and the two sets of qubits in between constitute the two branches. If no error occurs during the split repetition encoding, the state after $U_{\rm{encoding}}$ reads $(a\ket{0}\ket{0\hdots 0}\ket{0\hdots 0}+b\ket{1}\ket{1\hdots 1}\ket{1\hdots 1})\ket{0}$. Indeed, the last CNOTs targeting the flag qubit from each branch leave it in $\ket{0}$, since both controls are in the same state. However, if one qubit flips on one of the branches, say the first one, the state becomes $(a\ket{0}\ket{0\hdots 0}\ket{0\hdots 1}+b\ket{1}\ket{1\hdots 0}\ket{1\hdots 1})\ket{1}$. Therefore, a value $1$ observed of the flag qubit detects an error during the encoding. In such case, the associated outcome bit string will be discarded. On the other hand, results with the flag qubit having value $0$ are kept, and a majority vote is performed on the other qubits as usual. Notice that, despite an improved resilience to encoding errors with respect to chain and split encoding, there are still several cases in which the procedure can fail, for example if more than one error occur during the encoding, or if the readout of the flag qubit is faulty. In the following, we will investigate how the three repetition encoding types (chain, split and circular) compare in terms of robustness and under which conditions they can be used to effectively reduce readout error rates.

\subsection{Modelling hardware noise}
\label{sec:analytic_noise_model}

To support the analysis of different repetition encodings and to understand their dependence of various error sources, we introduce a theoretical noise model to which we will compare experimental results. 

The model is designed in order to keep the complexity and the number of parameters at a minimum, while still capturing the essential features affecting the performances of the proposed strategy. In particular, we assume that the only significant sources of errors are the ones affecting two-qubit (CNOT) and readout operations, and that they are all uncorrelated. We use a depolarizing channel $\mathcal{N}_{CNOT}:\rho \mapsto (1-\frac{4}{3}p_{ij})U_{CNOT}\rho U_{CNOT}^{\dagger} + \frac{p_{ij}}{3}\mathbb{1}_4$ to model the noise affecting a CNOT gate between qubits $i$ and $j$ on a state $\rho$, where $\rho$ is the density matrix of the state, $U_{CNOT}$ the CNOT-unitary and $\mathbb{1}_4$ the 4 by 4 identity matrix.
By parameterizing the depolarizing channel in this way, the CNOT operation fails with probability $p_{ij}$ when acting on a basis state, e.g. $\mathcal{N}_{CNOT}(\ketbra{10}) = (1-p_{ij})\ketbra{11} + \frac{p}{3}(\ketbra{00} + \ketbra{01} + \ketbra{10})$.

Readout errors behave as bit-flip errors with probabilities $p_0^r$ and $p_1^r$ describing the probability to measure an outcome $1$ (respectively $0$) from state $\ket{0}$ (respectively $\ket{1}$). Readout errors are assumed equal for all qubits in the register.

Let us now consider the unencoded computational basis states $\ket{0}\ket{0\hdots 0}$ and $\ket{1}\ket{0\hdots 0}$, where as usual the first qubit corresponds to the root and the other $n_{rep}$ auxiliary qubits are used for the encoding. The repetition code strategy would, without noise, proceed exactly through the quantum states $\ket{0}\ket{0\hdots 0}$ and $\ket{1}\ket{1\hdots 1}$, yielding the measurement outcomes $(0,\hdots,0)$ and $(1, \hdots,1)$ respectively. However, in the presence of errors affecting the encoding operations, all $2^{n_{rep}+1}$ possible bit strings have in principle a non-zero probability of occurring at the readout stage. In particular, such a probability will be a function of the noise model parameters $p_{ij}\,\forall \langle i,j \rangle$, $p_{0}^r$ and $p_{1}^r$. Here we denote by $\langle i,j \rangle$ all pairs of qubits involved in a CNOT operation during the implementation of $U_{\rm{encoding}}$.

The majority vote strategy (see also Sec.~\ref{sec:repetition_codes}) assigns each measured bit string to a logical outcome 0 or 1. Therefore, we define the total probability of a given encoding strategy to fail as the cumulative probability of all the output bit strings whose decoded values do not correspond to the original computational basis state. In this way, we derive from the noise model a logical error function $P_{\ket{x}}(\{p_{ij}\}, p_{0}^r, p_{1}^r)$ representing the failure probability of encoding the state $\ket{x}$ in terms of the CNOT and readout error rates.

\paragraph*{Two-qubit gate error inhomogeneity.} A possible additional simplification of the theoretical noise model presented above could be obtained by replacing all the individual CNOT error probabilities $p_{ij}$ by a single average value. However, such a drastic assumption would totally overlook the fact that, due to error-propagation mechanisms, a single CNOT error can have very different effects on the final result depending on the position at which it occurs. As already noticed above, for example, in the chain layout a high error probability on one of the very first links will lead to a significantly higher logical error rate compared to a similar situation affecting only the CNOT between the last pair of qubits. In view of the significant degree of variability among CNOT fidelities on real processors, we will therefore treat all $p_{ij}$ as independent random variables drawn from identical gaussian distributions $p_{ij} \sim \mathcal{N}(\mu = \bar{p}_{CNOT}, \sigma = \sigma_{CNOT})$ (truncated to remain in the interval $[0,1]$). The set of model parameters $\{p_{ij}\}$ is therefore reduced to the mean $\bar{p}_{CNOT}$ and the standard deviation $\sigma_{CNOT}$ of such distribution.

\paragraph*{Readout error asymmetry.} 
In the following, we will also make the approximation $P_{\ket{x}}(\bar{p}_{CNOT}, \sigma_{CNOT}, p_{0}^r, p_{1}^r) \approx P(\bar{p}_{CNOT}, \sigma_{CNOT}, p^r)$, namely we will neglect the effect of the asymmetry between the readout bit-flip errors for states $\ket{0}$ and $\ket{1}$ on the logical error function. The validity of this assumption, which greatly simplifies the forthcoming analysis, will be assessed a posteriori when comparing the predictions of the noise model with actual experimental data. In particular, we will always study the logical error function as a function of $p^r$ and clearly distinguish the results for the encoding of $\ket{0}$ and $\ket{1}$, which will often be characterized by different average readout errors. Moreover, it is worth noticing that when encoding either of the two logical states, the readout error of the other only occurs if a qubit was already flipped by an erroneous CNOT-gate. 
Thus, in practice, the readout error of the state that is not encoded, and therefore the readout error asymmetry itself, is suppressed by $\bar{p}_{CNOT}$. A more detailed analysis of the effect of readout asymmetry on the noise model is presented in the Supplementary Information, where we essentially confirm that its impact is negligible within the relevant range of parameters.

As a result of both simplifications regarding the CNOT and readout error probabilities, it turns out that the logical error function defined above is independent of the state $\ket{x}$ that is encoded and it is well approximated by a distribution depending on 3 parameters only, i.e.~$P_{\ket{x}}(\{p_{ij}\}, p_{0}^r, p_{1}^r)\approx P(\bar{p}_{CNOT}, \sigma_{CNOT}, p^r)$.
In the plots depicting this distribution that we will present in the following sections, the logical error will be shown as a function of $p^r$, for fixed values $\bar{p}_{CNOT}$ and $\sigma_{CNOT}$. 
The latter will be chosen to match as closely as possible the typical distribution of CNOT errors on the real devices employed for hardware tests. 
For a fixed readout error $p^r$, we compute the corresponding logical error by sampling the values of the $p_{ij}$s from the gaussian CNOT error distributions and applying the usual rules for computing the probability of independent events.
The resulting distribution is close to a gaussian 
and we use its average and $2\sigma$-interval to represent the logical error with its confidence interval in all subsequent $p^r-P$ plots.
For further details and a demonstrative example for the 2+1 chain repetition we refer also to Appendix~\ref{app:2p1_theoreticalnoisemodel}.

\section{Results} 
\label{sec:results}

\subsection{Benchmarking repetition codes}
\label{sec:benchmarking_rep_codes}

\begin{figure*}[t]
    \centering
    \begin{tikzpicture}
\node[anchor=north] (A1) at  (0,0) {
    \includegraphics[width=0.98\linewidth]{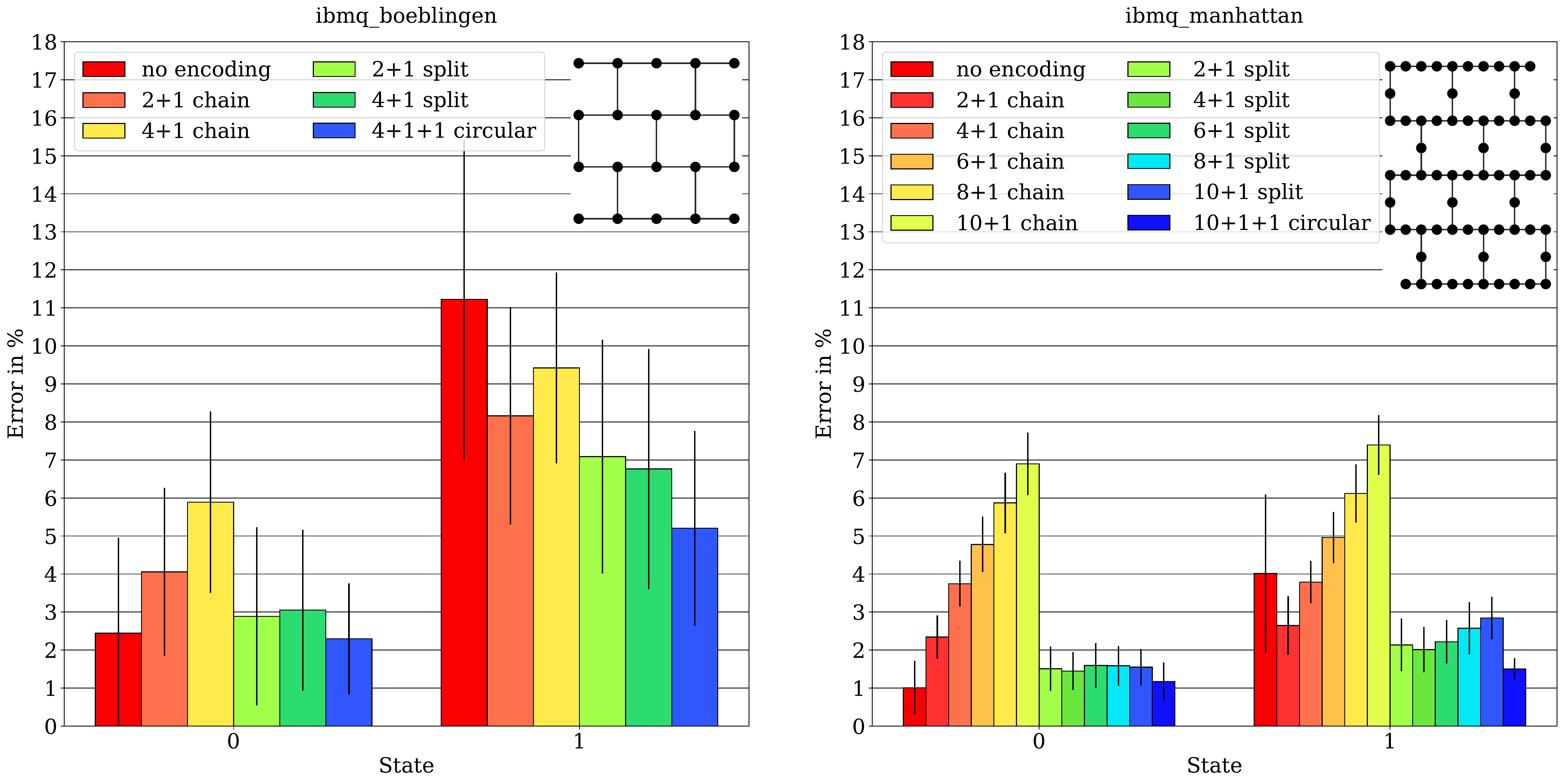}
};

\node[text] at (-8.5,-0.3) {(a)};
\node[text] at (0.5,-0.3) {(b)};
 \end{tikzpicture}
    \caption{Logical error rates of (un)encoded $\ket{0}$- and $\ket{1}$-states, averaged over different qubit layouts. 
    Vertical black line at each bar shows the standard deviation for each experiment on the qubit subset of (a)\textit{ibmq\_boeblingen} and (b)\textit{ibmq\_manhattan}.  Both in (a) and (b) we provide a schematic picture of the hardware used. For details on the hardware connectivity and the qubits used for each experiment see also Appendix \ref{app:ibmq_hardware}.}
    \label{fig:benchmark_all_encodings_hardware}
\end{figure*}

We assess the performances of the different repetition encoding schemes outlined above by analyzing their logical error rates on the single qubit basis states $\ket{0}$ and $\ket{1}$. 
For our analysis we name the different repetition encodings according to the repetition, flag and physical qubits as $n_{\rm{rep}} + n_{\rm{phys}}$ for the linear and chain encoding and as $n_{\rm{rep}} + n_{\rm{flag}} + n_{\rm{phys}}$ for the circular encoding. 
In Fig.~\ref{fig:benchmark_all_encodings_hardware}a, we report the average readout errors obtained on the IBM Quantum superconducting processor \textit{ibmq\_boeblingen} for using $\{2,4\}+1$-qubit chain and split repetition codes, as well as the $4+1+1$-qubit circular encoding.
In Fig.~\ref{fig:benchmark_all_encodings_hardware}b, we repeat the same calculation for the
\textit{ibmq\_manhattan} processor, using $\{2,4,6,8,10\}+1$ qubits for the chain and split encodings and the $10+1+1$-qubit circular encoding, that corresponds to the  heavy-hexagonal native architecture. 
The outcomes are compared with the corresponding unencoded case, in which the single qubit basis state is prepared and measured without performing any error mitigation.

We first notice a strong asymmetry between the unencoded states $\ket{0}$ and $\ket{1}$, the latter being significantly more error prone. 
It is therefore not surprising to observe that the proposed error mitigation strategy is particularly helpful in this second case. 
The effect of error propagation is also clearly visible from the results of the chain encodings, whose performances become worse the more qubits are used (approximately~$\sim1\%$ additional error every two qubits). 
In fact, apart from the $3$-qubit ($n_{rep}=2$)
version for which there is at least a reduction of the readout asymmetry, the chain repetition codes are always worse than unencoded counterparts. 
At the same time, the split repetition codes show, as expected, significantly less dependence on the size of the encoding registers. This strategy already leads to a significant reduction of the readout error on $\ket{1}$.

The best performances are obtained, in this set of experiments, with the $11$-qubit circular encoding scheme. 
Indeed, the readout error on $\ket{1}$ is decreased by $\sim2.5\%$ with respect to the unencoded case, while the net error on $\ket{0}$ is comparable to single-qubit results. 
This behaviour, which ultimately reflects the balance between the benefits and the complexity overhead introduced by repetition encodings, confirms that little improvement is to be expected in the regime of a relatively high CNOT to readout error ratio (see also following discussion in Sec.~\ref{sec:comparison_analytic}). 
On a more technical side, it is also worth noticing that the differences between the $11$-qubit split and $11$-qubit circular repetition codes denote the effect of the additional flag qubit.

\begin{figure}[ht!]
    \centering
\begin{tikzpicture}
\node[anchor=north] (A1) at  (1.7,0) {
    \includegraphics[width=0.98\linewidth]{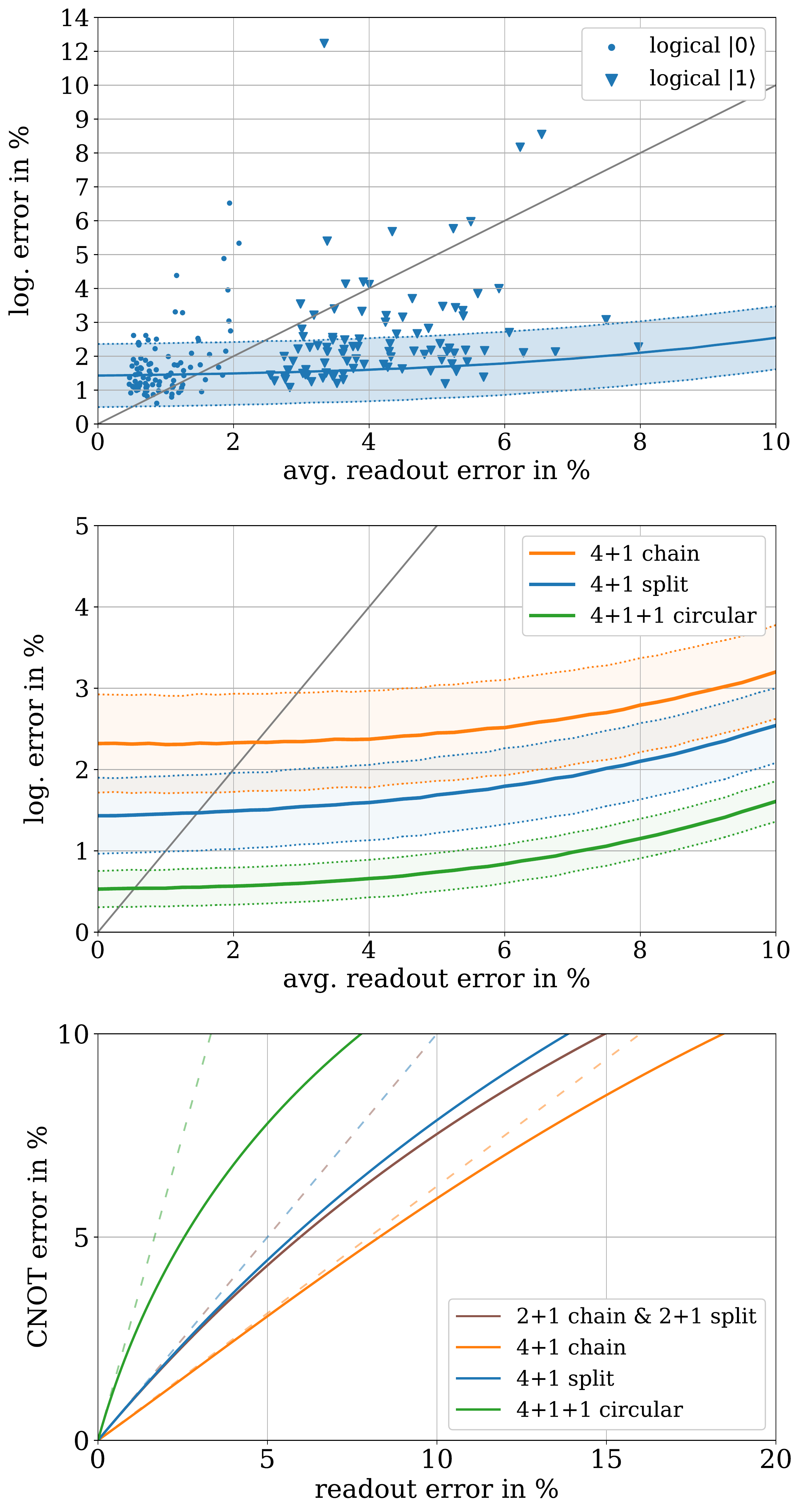}
    };

\node[text] at (-2.3,-0.3) {(a)};
\node[text] at (-2.3,-5.7) {(b)};
\node[text] at (-2.3,-11.1) {(c)};

    \end{tikzpicture}
\caption{(a) The logical error of the 5-qubit split encoding as a function of the average readout error, obtained by experiment (scattered points) and as predicted by the theoretical noise model (blue band). 
Dotted markers correspond to the logical $\ket{0}$ errors and triangles to the logical $\ket{1}$ ones.
(b) Logical error as a function of the average readout error for the $4+1$-qubit chain and split encoding and the $4+1+1$-qubit circular encoding, obtained by the theoretical noise model. (c) The crossover-curves show for which CNOT error the logical readout error of an encoding equals the average single qubit readout error. The dashed lines describe the asymptotic linear behaviour of these curves towards zero. 
}
\label{fig:Crossover-curve}
   
\end{figure}
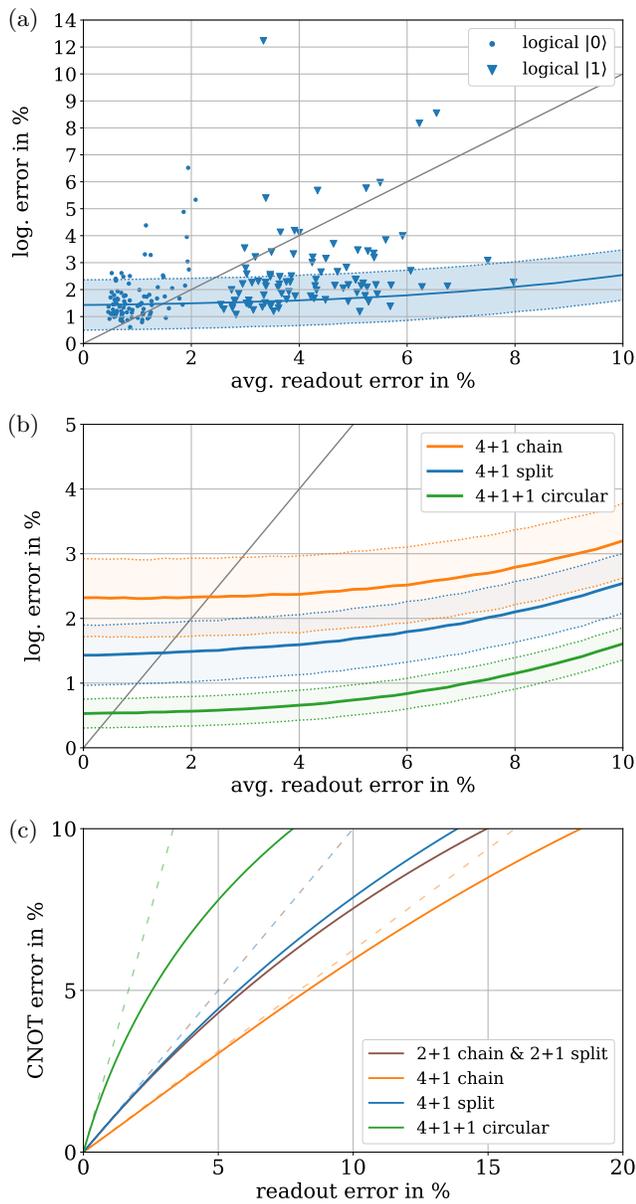

\begin{figure*}[ht!]
    \centering
\begin{tikzpicture}
\node[anchor=north] (A1) at  (1.2,0) {
\Qcircuit @C=0.75em @R=0.55em {
&q_0 ({\color{gray} \bf 33}) \ket{0} & \quad & \quad & \quad  & \qw & \qw  & \gate{U} & \qw & \qw  &\meter \\
}
};

\node[anchor=north] (A1) at  (1.2,-0.65) {
\Qcircuit @C=0.65em @R=0.08em {
&f ({\color{blue} \bf 47}) \ket{0}  & \quad & \quad &\quad  & \qw        & \qw      & \qw       &\qw        &\qw       &\qw       &\qw       &\targ     &\targ      &\meter \\
& & & \\ 
&a_5 ({\color{red} \bf 46})\ket{0} & \quad & \quad &\quad   & \qw        & \qw      & \qw       &\qw        &\qw       &\qw       &\targ     &\ctrl{-2} &\qw        &\meter \\
&a_4 ({\color{red} \bf 45})\ket{0} & \quad & \quad &\quad   & \qw        & \qw      & \qw       &\qw        &\qw       &\targ     &\ctrl{-1} &\qw       &\qw        &\meter \\
&a_3 ({\color{red} \bf 39})\ket{0} & \quad & \quad &\quad   & \qw        & \qw      & \qw       &\qw        &\targ     &\ctrl{-1} &\qw       &\qw       &\qw        &\meter \\
&a_2 ({\color{red} \bf 31})\ket{0} & \quad & \quad &\quad   & \qw        & \qw      & \qw       &\targ      &\ctrl{-1} &\qw       &\qw       &\qw       &\qw        &\meter \\
&a_1 ({\color{red} \bf 32})\ket{0} & \quad & \quad &\quad   & \qw        & \qw      & \targ     &\ctrl{-1}  &\qw       &\qw       &\qw       &\qw       &\qw        &\meter \\
& & & \\ 
&q_0 ({\color{gray} \bf 33})\ket{0} & \quad & \quad &\quad  & \gate{U} & \ctrl{2} & \ctrl{-2} &\qw        &\qw       &\qw       &\qw       &\qw       &\qw        &\meter \\
& &  & \\ 
&b_1 ({\color{red} \bf 34})\ket{0} & \quad & \quad &\quad   & \qw        & \targ    & \qw       &\ctrl{1}   &\qw       &\qw       &\qw       &\qw       &\qw        &\meter \\
&b_2 ({\color{red} \bf 35})\ket{0} & \quad & \quad &\quad   & \qw        & \qw      & \qw       &\targ      &\ctrl{1}  &\qw       &\qw       &\qw       &\qw        &\meter \\
&b_3 ({\color{red} \bf 40})\ket{0} & \quad & \quad &\quad   & \qw        & \qw      & \qw       &\qw        &\targ     &\ctrl{1}  &\qw       &\qw       &\qw        &\meter \\
&b_4 ({\color{red} \bf 49})\ket{0} & \quad & \quad &\quad   & \qw        & \qw      & \qw       &\qw        &\qw       &\targ     &\ctrl{1}  &\qw       &\qw        &\meter \\
&b_5 ({\color{red} \bf 48})\ket{0} & \quad & \quad &\quad   & \qw        & \qw      & \qw       &\qw        &\qw       &\qw       &\targ     &\qw       &\ctrl{-14} &\meter \\
}};

\node[anchor=north] (A1) at  (9.7,-0.1) {
    \includegraphics[width=0.59\linewidth]{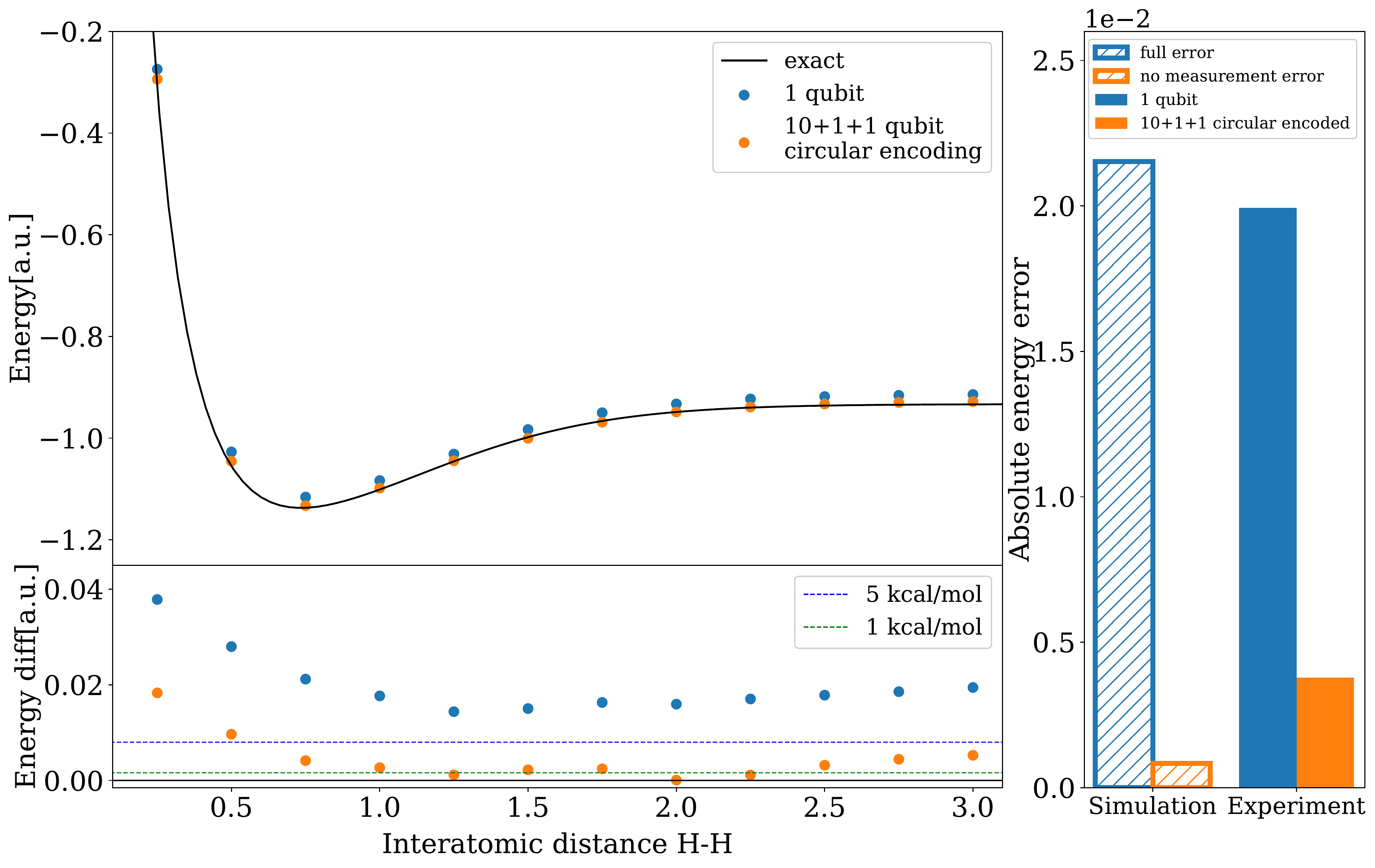}
};

\node[anchor=north] (A1) at  (7.75,-0.38) {
    \includegraphics[width=0.2\linewidth]{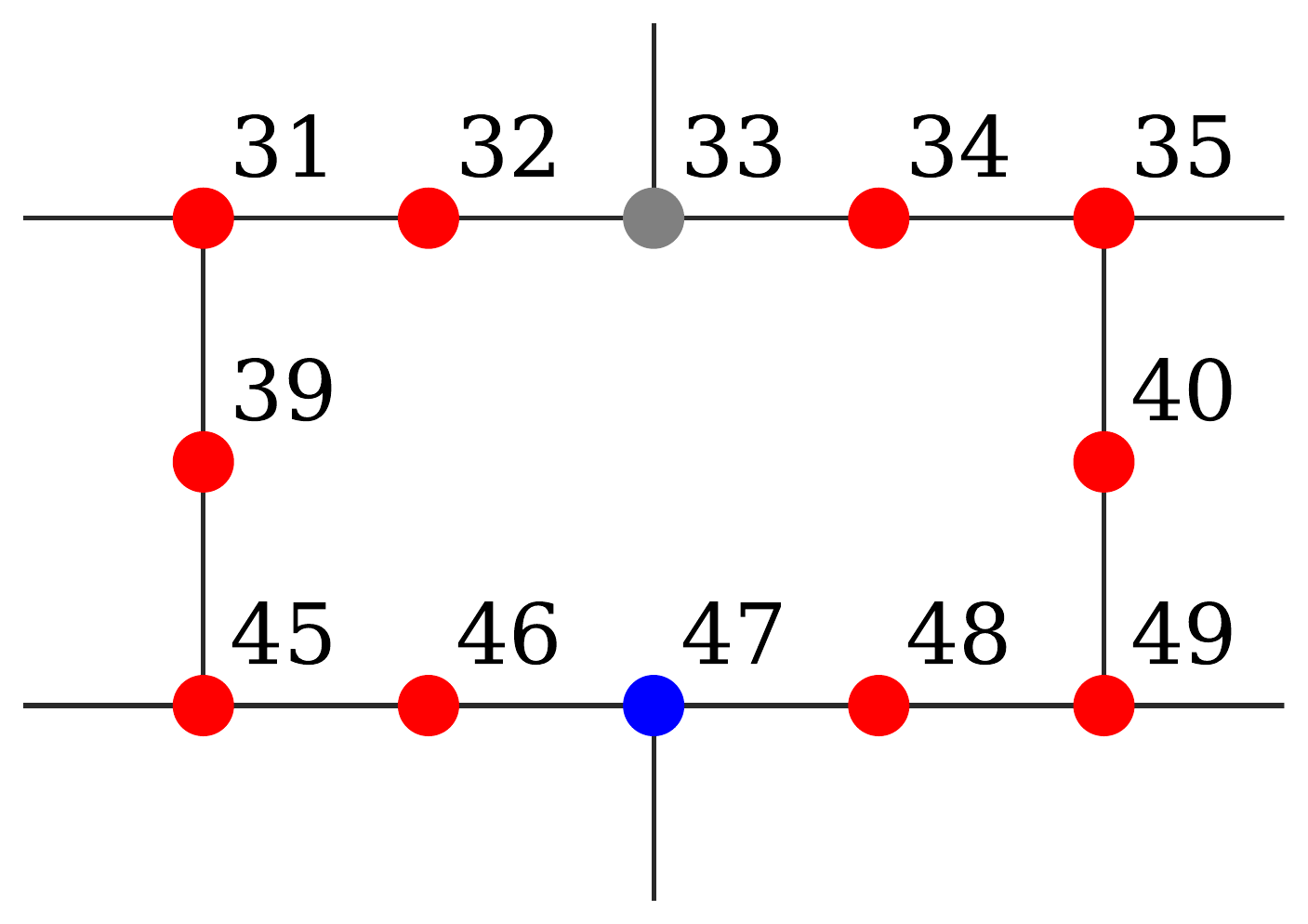}
};

\node[text] at (-2.5,-0.3) {(a)};
\node[text] at (-2.5,-0.78) {(b)};
\node[text] at (4.5,-0.3) {(c)};
\node[text] at (6,-0.7) {(d)};
\node[text] at (12.5,-0.3) {(e)};
 \end{tikzpicture}
    \caption{
    (a) Variational form used for the simuation of $\rm{H_2}$ molecule. Gate $U$ in our case is a parameterized $R_y$ rotation as in Eq.~\ref{eq:Ry_gate}.
    (b) The circuit after the application of the circular encoding. The qubits of the hardware used are highlighted in the parenthesis and the color code matches the one of panel (d). 
    (c) (upper) Energy expectation values of the H2 molecule at various interatomic distances. 
    Preoptimized parameters of no-noise simulations were used for preparing the states and measuring the expectation value.
    (lower) Energy difference with the exact solution. The blue dotted line corresponds to accuracy of 5 kcal/mol and the green dotted line to 1 kcal/mol (chemical accuracy). 
    (d) Layout of $ibmq\_manhattan$ hardware used for the simulation of $\rm{H_2}$ molecule. 
    The color code is red for repetition auxiliary qubits, gray for the physical qubits and blue for the flag qubits.
    (e) Total error for the circuit encoding in simulation and real hardware calculations averaging the results along the PES. The simulation results are done with the full error noise model of the corresponding hardware and the same noise model without considering the measurement errors.}
    \label{fig:H2_results_and_circuits}
\end{figure*}

\subsection{Comparison with the theoretical noise model}
\label{sec:comparison_analytic}

In this section, we discuss how logical readout error rates depend on the physical CNOT and readout errors with the help of the 
noise model introduced in Sec.~\ref{sec:analytic_noise_model}. 
In particular, we compare the $4+1$-qubit chain and split and $4+1+1$ circular repetition encoding to each other and to the theoretical predictions.

Figure~\ref{fig:Crossover-curve}a shows readout errors for the encoded states $\ket{0}$, $\ket{1}$ using the split repetition encoding on $96$ different $4+1$-qubit subsets on the \textit{ibmq\_manhattan} quantum processor. 
The mean and standard deviation of the distribution of CNOT errors across the entire device, obtained from calibration data provided within the IBM Quantum hardware Qiskit providers~\cite{Qiskit_short},
are used to construct the theoretical noise model. 
At the same time, the physical readout errors for individual qubits are determined empirically by preparing each qubit in $\ket{0}$ or $\ket{1}$ and performing measurements in the computational basis. 
The logical readout error obtained on each $4+1$-qubit subset is reported as a function of the corresponding average value $\bar{p}_{phys}^r$ for the physical single-qubit readout error measured on the constituents. 
Here, the readout error parameter $p^r$ is assumed equal for all registers and corresponds to the average readout error on all the qubits on the real hardware.

For all data points below the grey diagonal line in Fig.~\ref{fig:Crossover-curve}a, the encoding strategy leads to a reduction of the readout error with respect to the single-qubit unencoded case. 
Consistently with the benchmark cases reported in Sec.~\ref{sec:benchmarking_rep_codes}, we observe that the proposed method is particularly effective whenever the physical readout error is high. 
This happens more often for the basis state $\ket{1}$, whose average single qubit readout error typically lies between $2\%$ and $6\%$. 
However, there also exists a region towards $\bar{p}_{phys}^r \simeq 0$ in which the CNOT overhead introduced by the encoding procedure leads to worse overall performances with respect to unencoded states. 
A similar plot, comparing hardware data for the $4+1$-qubit chain repetition code to the noise model is provided in Appendix~\ref{app:theoretical_noise_model_comparison}.

Although the qualitative agreement between the data and the theoretical noise model is good, the latter does not fully capture all the features present in hardware results, mainly due to the many approximations involved in its construction (see Sec.~\ref{sec:analytic_noise_model}). In fact, numerical simulations of all the proposed repetition encodings performed with the software package Qiskit~\cite{Qiskit_short}, which allows one to emulate the real processors with full control over the noise processes, show high agreement with the predictions of the theoretical noise model. As a result, the theoretical noise model is to be understood essentially as an upper bound for the actual performances. We also notice that the hypothesis of a gaussian distribution for CNOT errors is not always fulfilled on real devices (see Appendix~\ref{app:cnot_error_distribution}, Fig.~\ref{fig:Cnot_err_distr_manhattan}).

To demonstrate the theoretical advantages brought by different repetition code layouts, we compare in Fig.~\ref{fig:Crossover-curve}b their logical error curve using the same underlying CNOT error distributions, with $\bar{p}_{CNOT} = 1.393\%$ and $\sigma_{CNOT} = 0.66\%$. 
All three error bands have very similar shapes, but they are shifted vertically highlighting the better performances of the split and circular approaches with respect to the chain layout. 
Indeed, a lower vertical position of the curve reduces the range of values $\bar{p}_{phys}^r \to 0$ for which the repetition code fails by design. 
A similar effect is also obtained, without changing the code layout, whenever $\bar{p}_{CNOT}$ decreases. 
We also notice that the confidence interval for the circular repetition code readout error is the smallest among the three different versions, denoting a higher resilience to error propagation effects.

As already mentioned, the crossover point between the logical readout error curve for a given encoding strategy and the diagonal line representing the unencoded error rate can be used as a proxy for the performances of said encoding. 
Fig.~\ref{fig:Crossover-curve}c shows such values in the $\bar{p}_{CNOT}$-$\bar{p}_{phys}^r$ plane.
For small error rates, these crossover curves can be approximated linear, as marked by dashed lines in the plot. The slope of these tangent curves equals the ratio $\bar{p}_{CNOT}/\bar{p}_{phys}^r$ at the crossover and therefore can be used, within the assumptions of the theoretical noise model, as a rule of thumb for assessing the usefulness of a given encoding. 
As an example, for the $4+1$-qubit chain repetition encoding, such condition turns out to be $\bar{p}_{CNOT} < (5/8)\bar{p}_{phys}^r$. Similarly, for the $2+1$- and $4+1$-qubit split encoding and the $3$-qubit chain encoding the threshold is $\bar{p}_{CNOT} < \bar{p}_{phys}^r$, while for the circular repetition encoding it is $3\bar{p}_{CNOT} < \bar{p}_{phys}^r$.

\subsection{Applications of circular encoding}

The results reported in the previous section, including experiments on real quantum processors, demonstrate that the circular repetition encoding has a good potential for achieving effective readout error mitigation. 
This is particularly true in view of the specific hardware characteristics (e.g. IBM Quantum hardware), whose heavy hexagonal connectivity can naturally host circular layouts. 
We will now provide some practical examples, implemented on real superconducting quantum processors, in which the proposed repetition encoding is successfully applied to typical problems of interest in chemistry and physics. 

\paragraph*{Electronic structure calculations} One promising near-term application of noisy quantum computers is the efficient calculation of electronic structure properties of chemical compounds. 
This is usually achieved with the use of adaptive quantum circuits: for example, the well known Variational Quantum Eigensolver (VQE) algorithm~\cite{Peruzzo2014, vqe1, vqe2} makes use of quantum resources to approximate electronic eigenstates and their corresponding energy expectation values $\expval{H}{\Psi(\theta)}$. 
Here, $H$ is the electronic molecular Hamiltonian describing the system under study and $\ket{\Psi(\theta)}$ is a parameterized wavefunction ansatz encoded on a qubit register. 

We investigate the effect of our proposed error mitigation strategy on a key ingredient of the VQE, namely the readout of energy expectation values, for the case of the dissociation profile of the H$_2$ molecule. 

In order to map the electronic structure Hamiltonian to a sum of Pauli operators we employ the parity mapping fermion-to-qubit transformation~\cite{taper}. 
The encoding of one orbital to one qubit lead to a 4 physical qubit system for the minimal STO3G basis set~\cite{sto3g}. 
Intrinsic properties of the parity mapping allow us to reduce the number of qubits by two~\cite{taper}, resulting in a 2-qubit Hamiltonian
\begin{equation}
   H^{2q}_{H_2} = h_0II+h_1IZ+h_2ZI +h_3ZZ+h_4XX,
\end{equation}
with $\ket{01}, \ket{10}$ as ground and excited states.
At this point with a basis change we encode the ground state of the system in the state $\ket{00}$ and the excited state in state $\ket{11}$ for the two physical qubits.
We further reduce the space by one more qubit due to system symmetries~\cite{taper}, leading to the single qubit problem
\begin{equation}
    H^{1q}_{H_2} = (h_0 - h_3)I + (h_2-h_1)Z + h_4X.
\end{equation}
For the parametrization of the circuit ansatz we use a single qubit $R_y(\theta)$ rotation, 
\begin{equation}
    R_y(\theta) = \begin{pmatrix}\cos(\theta/2) & -\sin(\theta/2) \\ \sin(\theta/2) & \cos(\theta/2)\end{pmatrix}.
\label{eq:Ry_gate}
\end{equation}
The corresponding circuit in both unencoded and circular encoding configurations are depicted in Fig.\ref{fig:H2_results_and_circuits}.

We perform a $10+1+1$-qubit quantum chemistry experiment for the $H_2$ molecule on the \textit{ibmq\_manhattan} processor, using the qubit layout depicted in Fig.~\ref{fig:H2_results_and_circuits}d. We use numerically optimized ansatz parameters to sample energy estimates beyond the reference results. Although the overall accuracy obtained on the device is already very good in the unencoded case, due to the simplicity of the quantum circuit (see Fig.\ref{fig:H2_results_and_circuits}a),
the effect of the circular repetition is clearly visible in Fig.~\ref{fig:H2_results_and_circuits}c. 
With the use of the circular encoding scheme we achieve accuracy below 5 kcal/mol for most points along the dissociation profile and for certain points close to equilibrium we reach the so called chemical accuracy (below 1 kcal/mol) within the chosen basis set. 
For any of the points the total error is decreased by effect of the encoding and lies below $20$ mHa in all cases. 
With a similar experiment, we also observe advantageous behaviour using a two qubit molecular system ($\rm{HeH^+}$, where we only measure the expectation value of a small perturbation to the reference Hartree-Fock state. 
More details can be found in Appendix \ref{app:UHF_HeH_calculation}).

\paragraph*{Digital quantum simulation of the Ising model} 

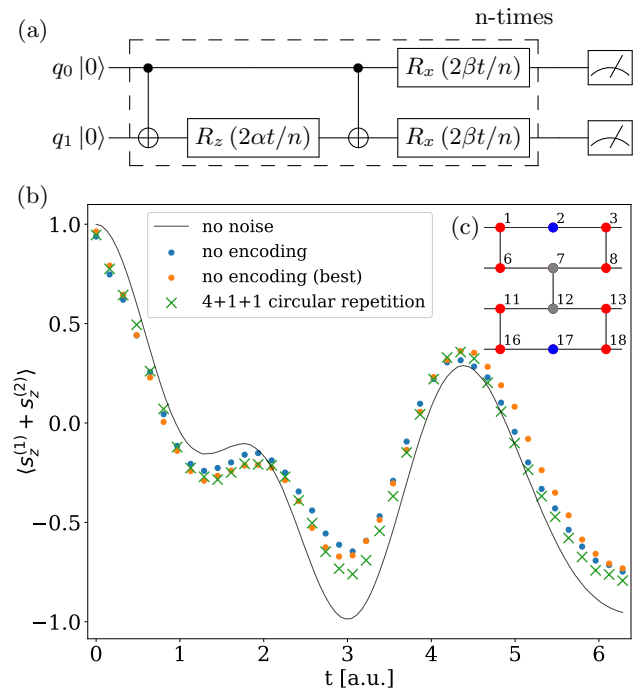
\begin{figure}[t!]
    \centering
\begin{tikzpicture}
\node[text] at (4.2,-1.3) { n-times };
\node[anchor=north] (A1) at  (1.9,-1.5) {
\Qcircuit @C=1.2em @R=1.3em {
&q_0 \ket{0} &\quad & \ctrl{1} & \qw & \ctrl{1} & \gate{R_x\left(2\beta t/n\right)} & \qw &\meter \\
&q_1 \ket{0} &\quad & \targ  & \gate{R_z\left(2\alpha t/n\right)} & \targ & \gate{R_x\left(2\beta t/n\right)} & \qw &\meter \gategroup{1}{4}{2}{7}{.7em}{--}\\
}

};

\node[anchor=north] (A1) at  (1.65,-3.6) {
    \includegraphics[width=0.97\linewidth]{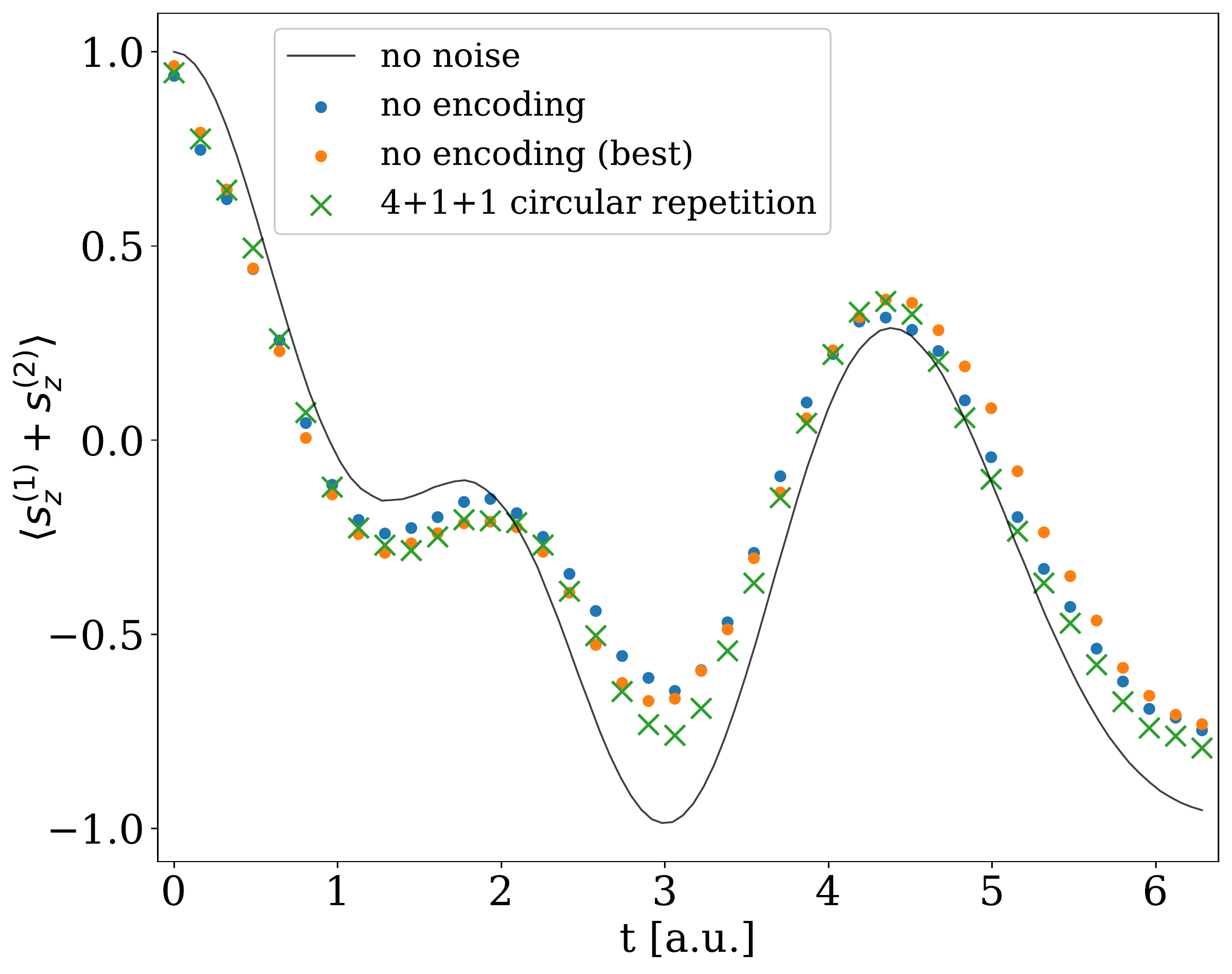}
};

\node[anchor=north] (A1) at  (4.75,-3.75) {
    \includegraphics[width=0.23\linewidth]{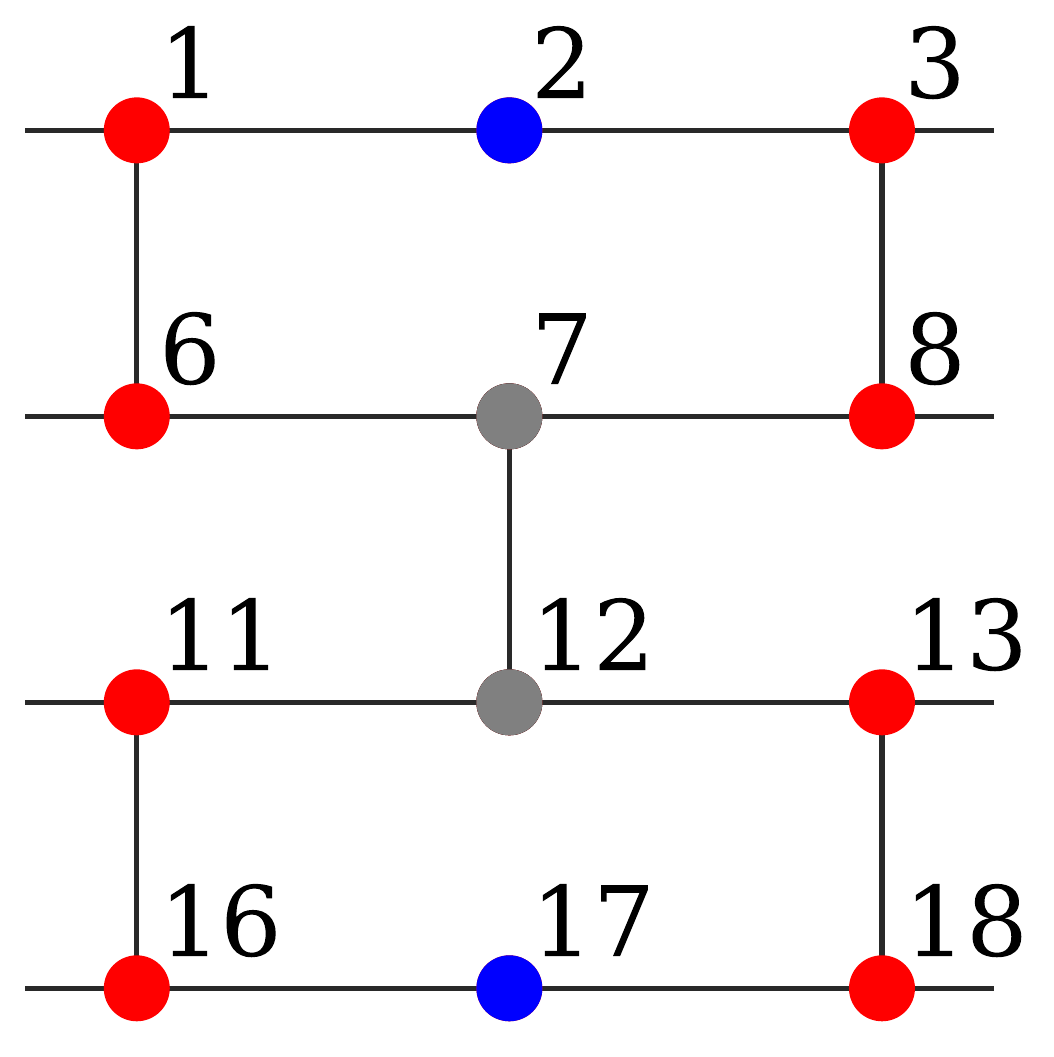}
};
\node[text] at (-2.2,-1.5) {(a)};
\node[text] at (-2.2,-3.7) {(b)};
\node[text] at ( 3.55,-4.12) {(c)};
\end{tikzpicture}
    \caption{(a) Circuit to perform the time evolution in the physical qubit space. 
    Each qubit is mapped using the circular encoding. 
    (b) Time evolution of the circuit with and without the circular encoding for the two physical qubits. 
    For the unencoded case we used qubits 7 and 12 for the general case (no encoding) and qubits 15 and 16 as the pair of qubits with minimum readout error(no encoding(best)). 
    The black solid line corresponds to the $5$-step Suzuki-Trotter approximation without noise and can be used for comparison.
    (c) Part of $ibmq\_singapore$ used for our calculations. The color code is red for repetition auxiliary qubits, gray for the physical qubits and blue for the flag qubits.}
    \label{fig:full_transverse_ising_5trotter}
\end{figure}

In order to illustrate the performances of the circular repetition encoding strategy on more challenging, deeper circuits, we implement the digital quantum simulation of the 2-spin transverse field Ising model~\cite{RevTacchino}
\begin{equation}
    H = \alpha Z_1 Z_2 + \beta (X_1 +X_2).
    \label{eq:Hamiltonian transverse Ising model}
\end{equation}
The real time propagator $\mathcal{U}(t) = e^{-iHt}$, which provides the dynamical evolution of the system, can be approximated by the Suzuki-Trotter product formula~\cite{trotter_product_1959,hatano_finding_2005,RevTacchino}
\begin{equation}
    \mathcal{U}(t) \approx \left(e^{-\frac{it}{n}\alpha Z_1 Z_2}e^{-\frac{it}{n}\beta(X_1+X_2)}\right)^n = \left[\,\mathcal{U}_n(t)\,\right]^n.
\end{equation}
Each Trotter step $\mathcal{U}_n(t)$ can be realized in practice with the combination of single-qubit rotations and CNOT gates reported in Fig.~\ref{fig:full_transverse_ising_5trotter}a~\cite{RevTacchino}.

The results for the implementation of the $n=5$ case on the \textit{ibmq\_singapore} quantum processor are shown in Fig.~\ref{fig:full_transverse_ising_5trotter}, where we depict the time evolution of the total spin component along the $z$ direction ($s_z = (\sigma_z^{(1)}+\sigma_z^{(2)})/2$). The pair of physical qubits of the quantum hardware used are qubits 7 and 12 (See also Fig.~\ref{fig:full_transverse_ising_5trotter}c and Appendix~\ref{app:ibmq_hardware}) 
was used both for the unencoded version and as root qubits in the $4+1+1$-circular repetition code. 
Furthermore, another pair of qubits (indices 15-16 on the device), reporting the lowest readout error calibration data, was also used for reference. 
As it can be seen, the circular encoding outperforms all the unencoded versions, including the best (in terms of readout error) available qubits of the device. 
It should be noted that the improvements due to readout error correction here are smaller compared to the electronic structure calculations.
This can be attributed to the fact that with the much longer circuits the errors caused by gate-operations are the main source of error.
Additional results, comparing the circular repetition code with the [4,2,2]-code for error detection, are provided in Appendix \ref{app:comparison_to_422}. 

\section{Conclusions}
\label{sec:conclusions}

In this work, we demonstrated how the repetition codes can be used as a starting point for quantum error correction in near term quantum computations. 
We analysed the effect of using different repetition encodings on IBM Quantum hardware and we provided a simplified, but rigorous, noise model that can predict the bounds of logical errors for our schemes. 
The different schemes scale linearly in terms of required qubits for the repetitions, while the introduced overhead of circuit depth is constant. The results are in good agreement with hardware calculations and provide insights to the limitations of the applicability of repetition encodings. 
We also demonstrated proof-of-principle applications of repetition encodings to quantum chemistry and quantum physics models, where we were able to retrieve more accurate solutions to the problems under consideration.

Our results show potential for the use of repetition codes for more accurate calculations, especially in view of future hardware developments. 
The proposed circular encoding strategy can be natively realized on qubit architectures like the heavy hexagonal lattice and does not lead to substantial overhead in terms of computational resources. 
The connectivity of encoded circular repetition blocks can possibly suffer from hardware limitations, but this effect can be mitigated using clever implementation strategies that need to be devised in parallel with the development of the hardware. 
The possibility to use other qubits from our encoding also to connect different repetition blocks is still to be explored and can also potentially mitigate the effect of restricted connectivity of the devices. 
Another important aspect is that our method is not limiting the integration with other error mitigation techniques and can actually be used with schemes like the zero noise extrapolation, without increasing the calculation overhead. 
We believe that this work can serve as a basis for future implementations and will provide clear guidelines for the use of repetition codes for applications in noisy near-term quantum computers.

\section{Acknowledgements}
JRW and IT acknowledge support from the Swiss National Science Foundation through NCCR SPIN. 
IBM, the IBM logo, and ibm.com are trademarks of International Business Machines Corp., registered in many jurisdictions worldwide. Other product and service names might be trademarks of IBM or
other companies. The current list of IBM trademarks is available at \url{https://www.ibm.com/legal/copytrade}.

\bibliography{Repetition_bib}
\bibliographystyle{unsrt}

\appendix

\section{Theoretical noise model demonstration}
\label{app:2p1_theoreticalnoisemodel}
Here we use a demonstrative example to illustrate the computation of the logical error function of the analytic noise model by going through the simple example of the 3-qubit chain repetition encoding. 
In particular, we consider here the encoding process of the $\ket{1}$ state, for which the pre-encoding state is $\ket{100}$, where the first qubit corresponds to the root and the three qubits are connected by CNOTs from left to right.

After the first CNOT, which is subject to a depolarizing error with probability $p_{0,1}$, the outcome state, if no error occurs, is $\ket{110}$ with probability $(1-p_{1,0})$. 
However the states $\ket{100}, \ket{000}$ and $\ket{010}$ are also a possible outcome state with an equal probability of $\frac{1}{3}p_{0,1}$. 

In the same way the application of the second CNOT leads to the following states with the corresponding probabilities:
\begin{align*}
    \ket{111}:& (1-p_{0,1})(1-p_{1,2}) + \frac{1}{9}p_{0,1}p_{1,2} \\
    \ket{110}:& (1-p_{0,1})\frac{1}{3}p_{1,2} + \frac{1}{9}p_{0,1}p_{1,2} \\
    \ket{101}:& (1-p_{0,1})\frac{1}{3}p_{1,2} + \frac{1}{9}p_{0,1}p_{1,2} \\
    \vdots \ \ \ \ \ &\\
    \ket{000}:& \frac{1}{3}p_{0,1}(1-p_{1,2}) + \frac{1}{9}p_{0,1}p_{1,2}
\end{align*}

Considering also the readout errors with error rate $p^r_{phys}$ each of the states can then result in either one of the $8$ physical states, so that their probabilities, including the CNOT error probabilities, will consist of $16$ terms, e.g. for $\ket{111}$:

\begin{align*}
    &\ket{111}:(p^r_{phys})^3\big(\frac{1}{9}p_{0,1}p_{1,2}+\frac{1}{3}p_{0,1}(1-p_{1,2})\big)\\
    &\ \ +4(p^r_{phys})^2(1-p^r_{phys})\frac{1}{9}p_{0,1}p_{1,2}\\
    &\ \ +(p^r_{phys})^2(1-p^r_{phys})\big(\frac{1}{3}p_{0,1}(1-p_{1,2}) + \frac{1}{3}(1-p_{0,1})p_{1,2}\big)\\
    &\ \ +(p^r_{phys})(1-(p^r_{phys}))^2\big(\frac{1}{9}p_{0,1}p_{1,2} + \frac{1}{3}p_{0,1}(1-p_{1,2})\big)\\
    &\ \ +2(p^r_{phys})(1-(p^r_{phys}))^2\big(\frac{1}{9}p_{0,1}p_{1,2} + \frac{1}{3}(1-p_{0,1})p_{1,2}\big)\\
    &\ \ +(1-(p^r_{phys}))^3\big(\frac{1}{9}p_{0,1}p_{1,2} + (1-p_{0,1})(1-p_{1,2})\big)
\end{align*}

After applying the majority vote, the sum of probabilities of all states that result in outcome $\ket{0}$ gives a logical error probability with the following form:

\begin{align*}
    &P(\{p_{0,1},p_{1,2}\}, p^r_{phys}) = \\ 
    &\ \ \ \ \frac{32}{9}(p^r_{phys})^3p_{0,1}p_{1,2} - \frac{8}{3}(p^r_{phys})^3p_{0,1}\\
    &\ \ \ \ - \frac{8}{3}(p^r_{phys})^3p_{1,2} + 2(p^r_{phys})^3 - \frac{16}{3}(p^r_{phys})^2p_{0,1}p_{1,2}\\
    &\ \ \ \ + 4(p^r_{phys})^2p_{0,1} + 4(p^r_{phys})^2p_{1,2} - 3(p^r_{phys})^2\\
    &\ \ \ \ + \frac{8}{9}(p^r_{phys})p_{0,1}p_{1,2} -\frac{2}{3}(p^r_{phys})p_{1,2} + \frac{4}{9}p_{0,1}p_{1,2}\\
    &\ \ \ \ - \frac{2}{3}p_{0,1} - \frac{1}{3}p_{1,2} + 1
\end{align*}

\section{Theoretical noise model validation}
\label{app:theoretical_noise_model_comparison}

\begin{figure}[ht!]
    \centering
    \includegraphics[width=0.95\linewidth]{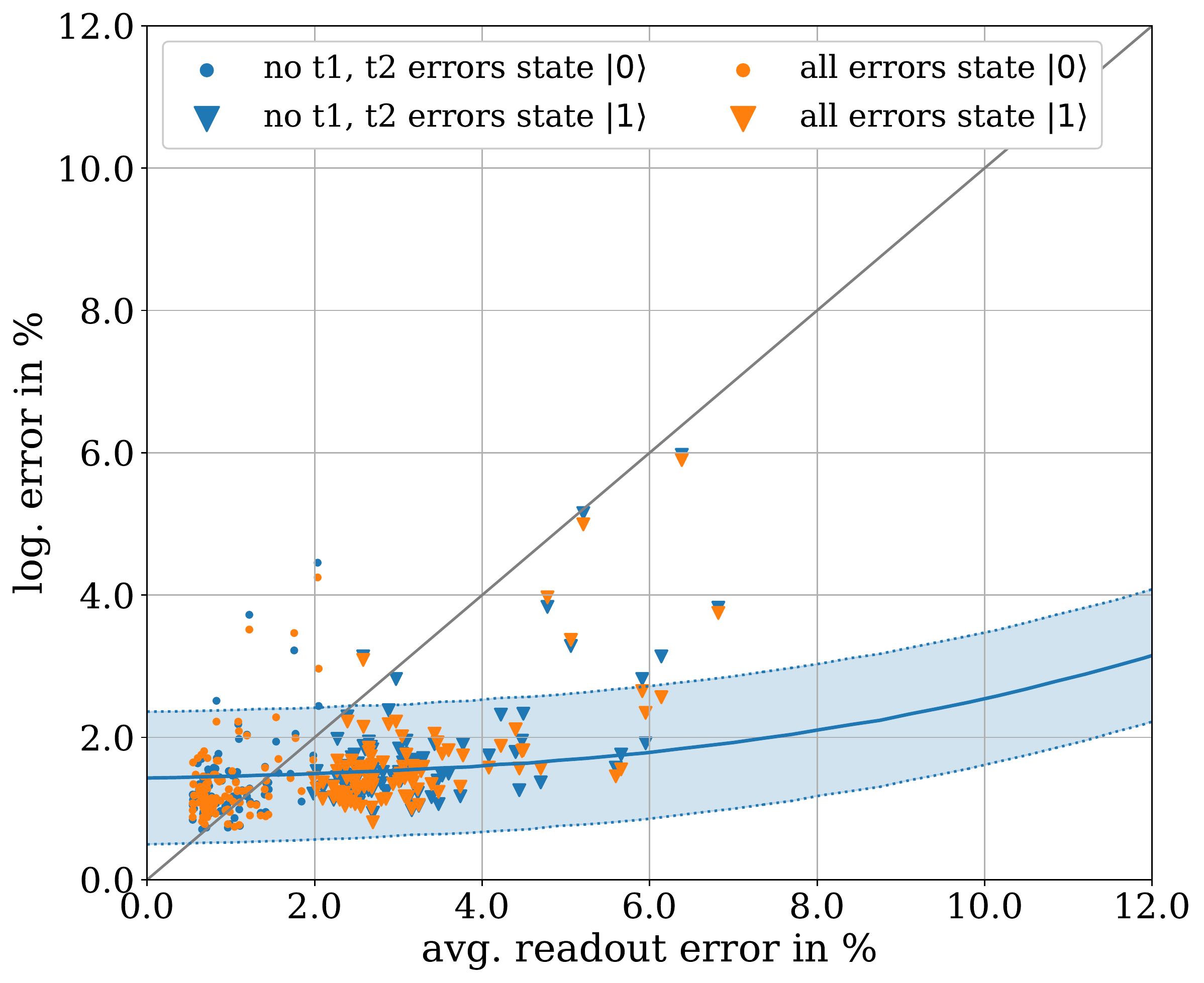}
    \caption{The logical error of the 5-split encoding as a function  of the  average readout error,  obtained by Qiskit simulation (scattered points) and as predicted by the theoretical noise model (blue band). Dotted markers correspond to the logical $\ket{0}$ errors and triangles to the logical $\ket{1}$ ones. In orange are data-points for which the full simulated hardware noise was used, for points in blue the errors caused by finite coherence times were ignored.}
    \label{fig:plot_5split_sim_manhattan_29_11_2020}
\end{figure}

\begin{figure}[ht!]
    \centering
    \includegraphics[width=0.95\linewidth]{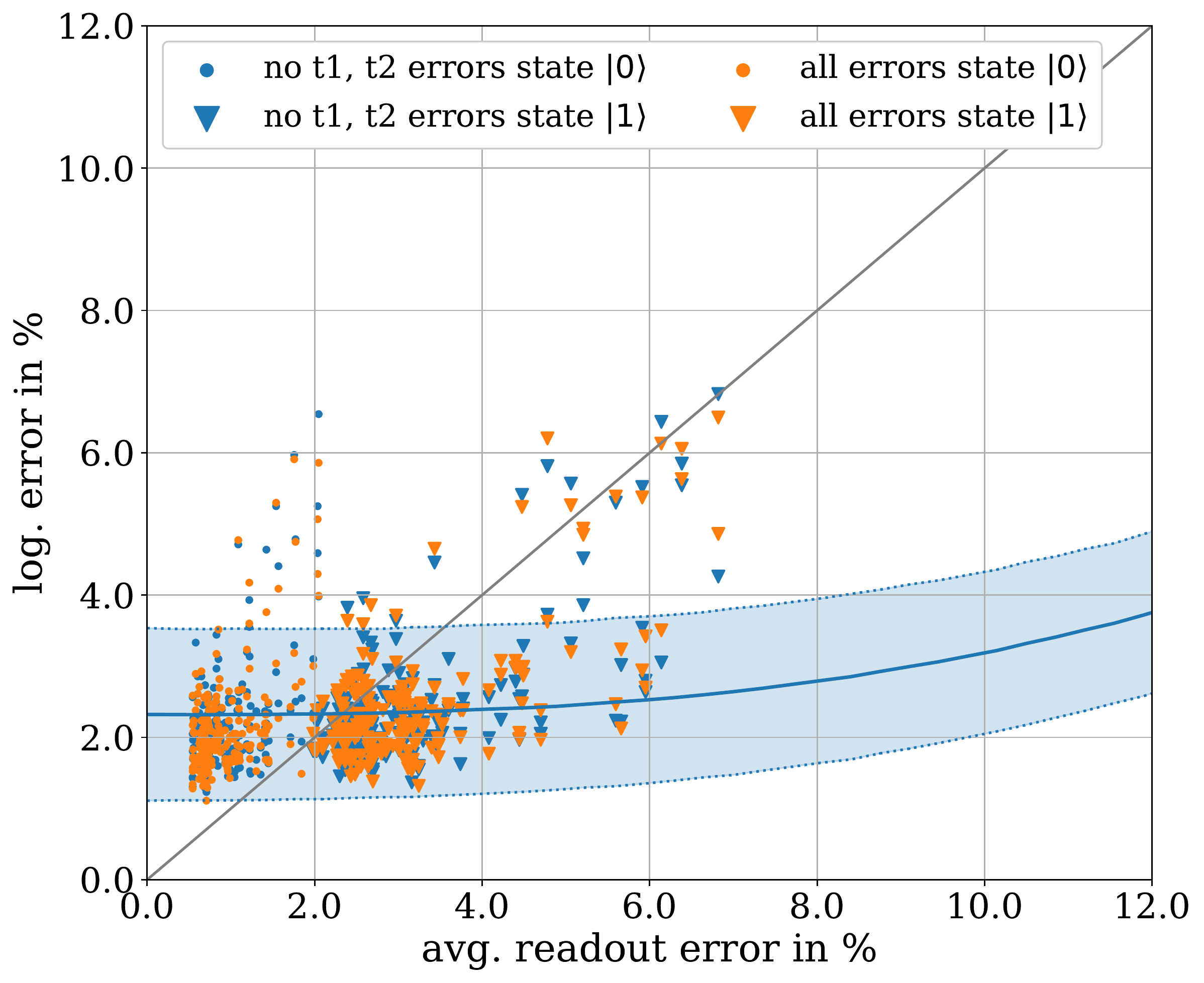}
    \caption{The logical error of the 5-chain encoding as a function  of the  average readout error,  obtained by Qiskit simulation (scattered points) and as predicted by the theoretical noise model (blue band). Dotted markers correspond to the logical $\ket{0}$ errors and triangles to the logical $\ket{1}$ ones. In orange are data-points for which the full simulated hardware noise was used, for points in blue the errors caused by finite coherence times were ignored.}
    \label{fig:plot_5chain_sim_manhattan_29_11_2020}
\end{figure}

In order to validate the proposed noise model, we benchmark it against noisy calculations in Qiskit~\cite{Qiskit_short} and quantum hardware experiments. 
Figure \ref{fig:plot_5chain_sim_manhattan_29_11_2020} and \ref{fig:plot_5split_sim_manhattan_29_11_2020} show, similarly to figure \ref{fig:Crossover-curve}a of the main text, the logical readout errors of encoded qubits as function of the average readout error $\bar{p}^r_{phys}$. Data points are obtained by noisy simulations using Qiskit realistic noise models simulations.
The plots highlight that the proposed theoretical noise model predicts the logical error of the vast majority of qubit subsets correctly. 
By comparing blue and orange points we notice, that neglecting the finite coherence time of qubits (t1, t2) is a reasonable approximation. 

\begin{figure}[ht!]
    \centering
    \includegraphics[width=0.95\linewidth]{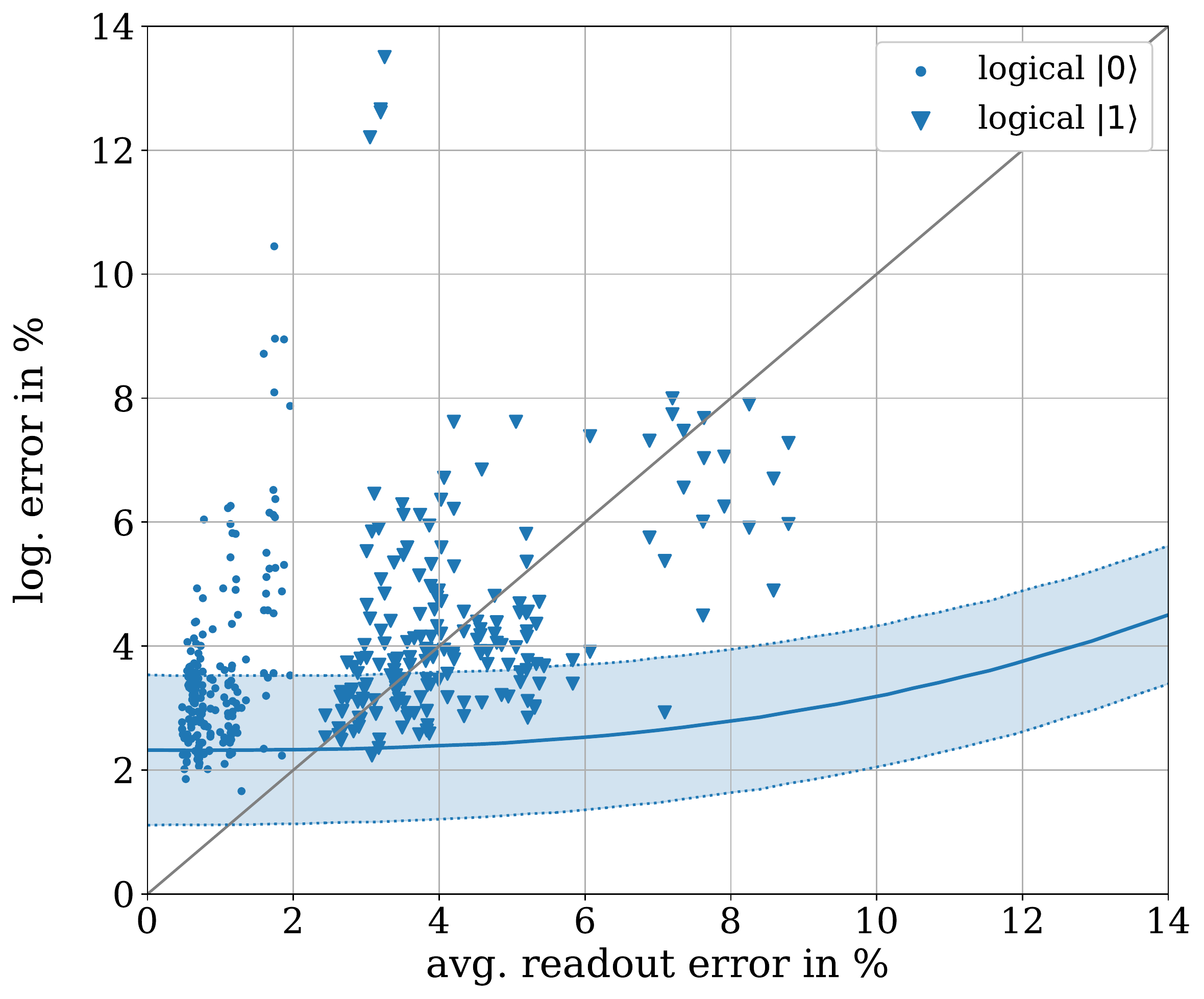}
    \caption{The logical error of the 5-qubit chain encoding as a function  of the  average readout error,  obtained by experiment (scattered points) and as predicted by the theoretical noise model (blue band). Dotted markers correspond to the logical $\ket{0}$ errors and triangles to the logical $\ket{1}$ ones.}
    \label{fig:plot_5chain_manhattan_29_11_2020}
\end{figure}

We also tested the model against \textit{ibmq\_manhattan} and the results show some agreement and confirmed that the noise model can provide a lower bound for the logical error estimates. 
For certain subset of qubits we notice that the noise model cannot describe properly the hardware behavior. 
Figure~\ref{fig:plot_5chain_manhattan_29_11_2020} demonstrates the hardware experiments and is the analog of Fig.~\ref{fig:Crossover-curve}a of the main text, but now applied for the chain repetition encoding.

\section{Applying repetition codes to the initial state of $\rm{HeH^+}$}
\label{app:UHF_HeH_calculation}

For the description of HeH$^{\rm{+}}$ we need 2 molecular orbitals corresponding to 4 spin orbitals (4 qubits).
For the simulation of the HeH$^{\rm{+}}$ we use a 2-qubit Hamiltonian arising from the 2-qubit reduction technique of parity mapping~\cite{taper}. 
The resulting Hamiltonians have the form 
\begin{align}
    \notag H_{HeH^{\rm{+}}} &= c_0 II + c_1 ZI + c_2 IZ + c_3 ZZ + c_4 XI \\ 
    &+ c_5 XZ + c_6 IX + c_7 ZX + c_8 XX
\end{align}
where the coefficients $c_i$ are dependent on the one- and two-body integrals calculated for every interatomic distance under consideration. 
The calculation of the coefficients is performed using the classical code PySCF~\cite{pyscf}, where we perform an Unrestricted Hartree Fock calculation (UHF).

To evaluate the performance of our scheme on a 2-qubit circuit, we prepare perturbations to the Hartree-Fock (HF) ground state determinant for HeH$^+$ with a $2$-qubit quantum circuit of the form 

\begin{equation*}
\Qcircuit @C=0.32em @R=0.32em {
& \gate{U_3\left(\frac{\pi}{2}-2a,\frac{-\pi}{2},\pi\right)} & \ctrl{1} & \gate{U_3\left(c,0,\frac{-\pi}{2}\right)} & \ctrl{1} & \gate{U_3\left(\frac{\pi}{2},0,\pi\right)} &\qw \\
& \gate{U_3\left(\frac{\pi}{2},2b-\frac{\pi}{2},\frac{\pi}{2}\right)} & \targ  & \gate{U_3\left(0,0,c\right)} & \targ &  \gate{U_3\left(\frac{\pi}{2},\frac{\pi}{2},\frac{-\pi}{2}\right)} & \qw \\
}
\end{equation*}

where $a$, $b$, $c$ are variational parameters and 
\begin{equation}
    U_3(\theta,\phi,\lambda) = \begin{pmatrix}\cos(\theta/2) & -e^{i\lambda}\sin(\theta/2) \\ e^{i\phi}\sin(\theta/2) & e^{i(\phi+\lambda)}\cos(\theta/2)\end{pmatrix}
\end{equation}
\begin{figure}[b]
    \centering
    \includegraphics[width=0.9\linewidth]{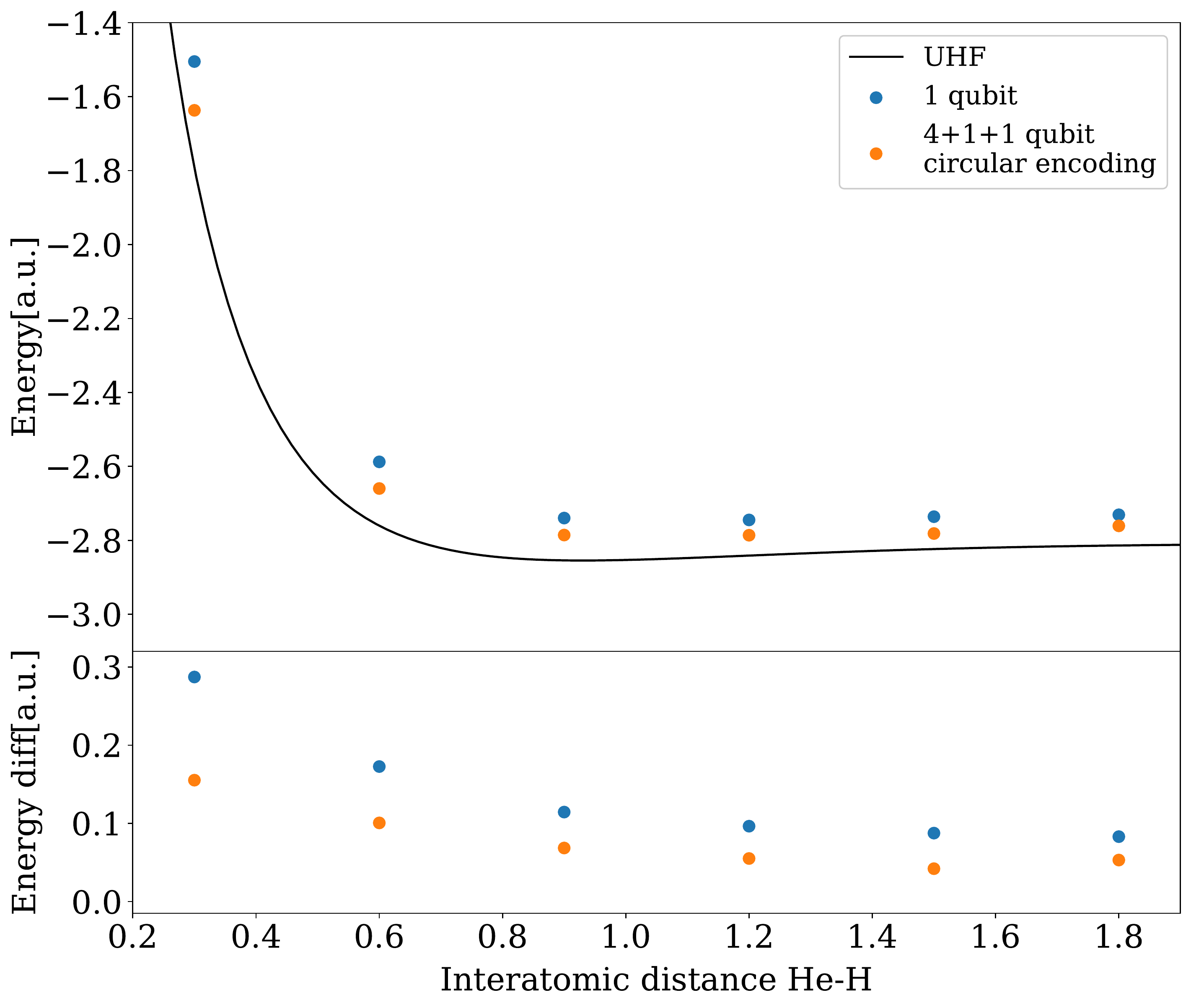}
    \caption{Application of repetition encoding for the calculation of the HF energy for the two qubit system HeH+ in \textit{ibmq\_singapore}.}
    \label{fig:singapore_HeH+_results}
\end{figure} 
Notice that this essentially corresponds to a UCCSD ansatz~\cite{Peruzzo2014, Barkoutsos2018}, which we use here as an approximation to the reference HF dissociation curve for different interatomic distances. 

The quantum circuit was implemented and run on the \textit{ibmq\_singapore} device, both with and without circular repetition encoding. 
For all data points, the same 2 qubits are used for the unencoded version and as root qubits for a $4+1+1$-qubit circular repetition code. 
In this way, the gate errors during the state preparation stage are consistent and the results only differ in the quality of the final measurements, highlighting the effect of the circular repetition encoding for readout error mitigation. 
In fact, as shown in Fig.~\ref{fig:singapore_HeH+_results}, the circular encoding leads to an average decrease of the total error (i.e.\ the difference from the reference energy values) by approximately ~$43\%$ with respect to the unencoded case. 
It is worth mentioning that, from numerical simulations done with Qiskit and based on the calibration data for the real device, it turns out that the fraction of such total error which can be ascribed to readout noise is typically around $50\%$, thus showing an almost complete cancellation due to the proposed circular encoding.

\section{Comparison to the [4,2,2]-code}
\label{app:comparison_to_422}
To demonstrate the effect of our scheme in comparison to small error detection codes we repeat the digital quantum simulation of the Ising model also the case of the [4,2,2]-error detection code. 

The [4,2,2] code encodes 2 logical qubits in 4 physical ones and is generated using the stabilizers $XXXX$ and $ZZZZ$. 
It can, in principle, detect any single-qubit error by stabilizer measurements, but due to the code-distance of only $2$ it can only detect and not correct for them. 
The three single-qubit rotations necessary to implement a single Trotter-step (see fig.~\ref{fig:full_transverse_ising_5trotter}~a) cannot be implemented transversally within the [4,2,2]-code, which means that each rotation cannot be performed by operations on a single physical qubit. 
In fact, six instead of three CNOTs are needed for each Trotter-step, highlighting the difficulty of implementing a circuit with an incomplete basis-set of logical operations. 
The limited qubit-connectivity of the hardware used for this experiment, \textit{ibmq\_singapore}, does not allow for stabilizer measurements. 
After the time-evolution circuit, the four physical qubits are measured in the computational basis and the results outside of the logical codespace are discarded.  The remaining 4-bit bitstrings are decoded back into their logical 2-qubit counterpart, from which the spin value $s_Z^{(1)}+s_Z^{(2)}$ can be evaluated.

\begin{figure}[ht!]
    \centering
    \includegraphics[width=0.86\linewidth]{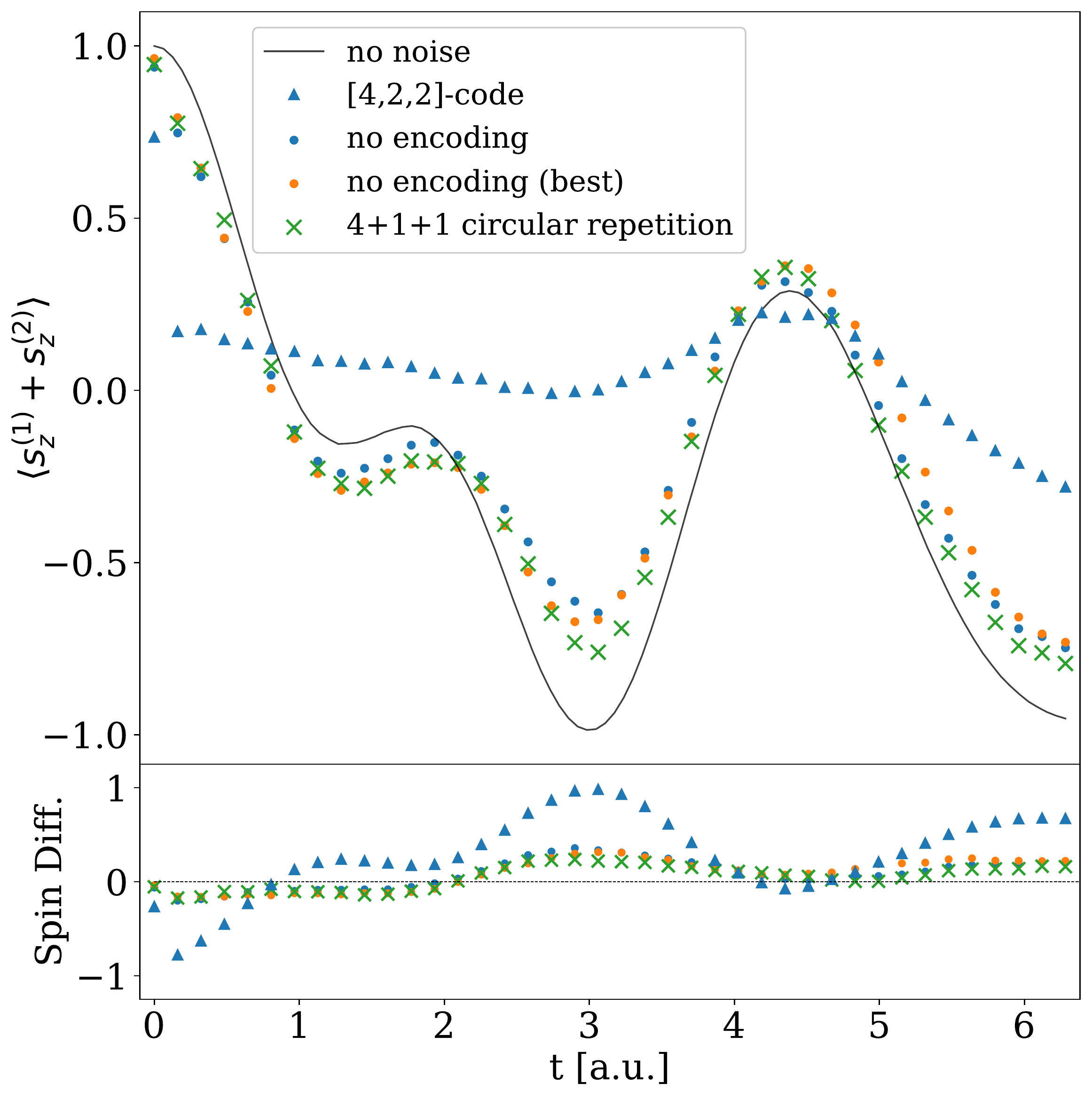}
    \caption{Time evolution of the transverse Ising-model with circular encoding, a [4,2,2]-encoding, and no encoding for the two physical qubits. 
    For the unencoded case we used qubits 7 and 12 for the general case (no encoding) and qubits 15 and 16 as the pair of qubits with minimum readout error(no encoding(best)). Qubits 0,1,2,6 of the \textit{ibmq\_singapore} device are used for the [4,2,2]-encoding. The black solid line corresponds to the $5$-step Suzuki-Trotter approximation without noise and can be used for comparison. In the lower panel the differences between the experimental data-points and the no-noise results are given.}
    \label{fig:422-comparison}
\end{figure}

\section{Readout asymmetry ratio}
\begin{figure}[ht!]
    \centering
    \includegraphics[width=\linewidth]{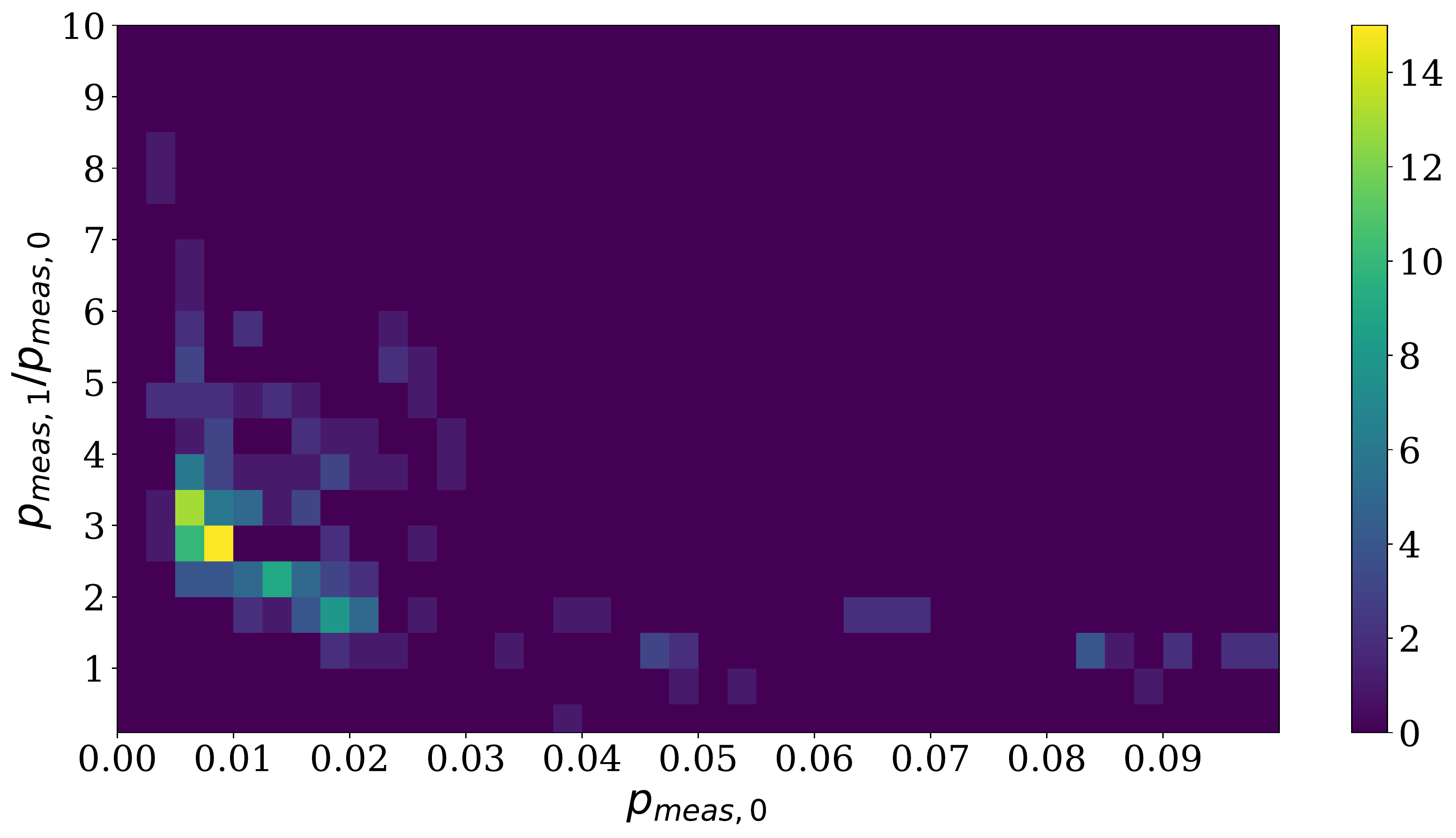}
    \caption{The asymmetry ratio $\frac{p_{meas,1}}{p_{meas,0}}$ as a function of $p_{meas,0}$ for all qubits on the $\textit{ibmq\_manhattan}$ device.}
    \label{fig:Asymmetry-ratio_distr}
\end{figure}

\begin{figure}[ht!]
    \centering
    \includegraphics[width=0.9\linewidth]{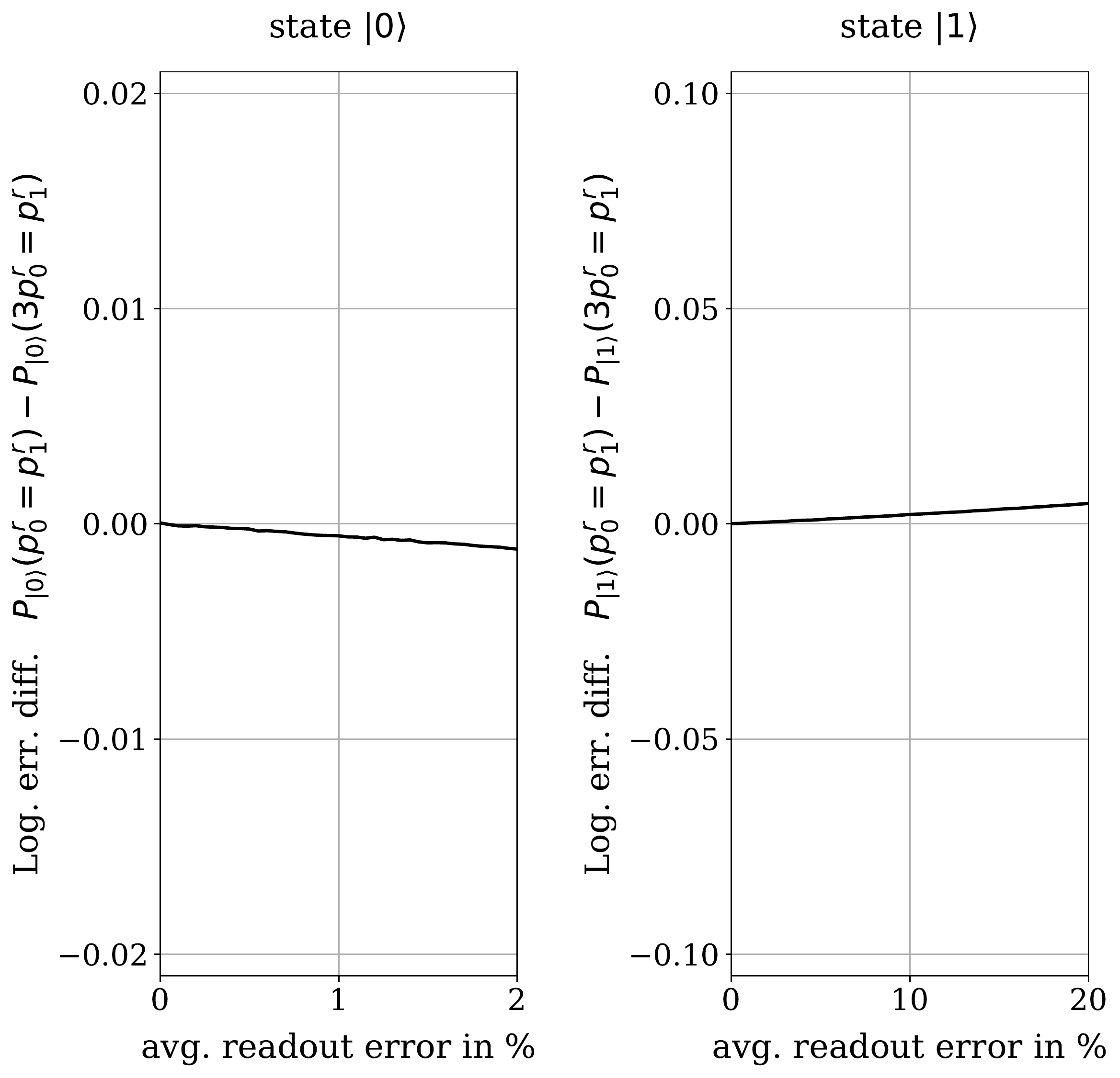}
    \caption{Error caused by the approximation $p^r_0=p^r_1$ for the $4+1$-qubit split code as a function of $p_{meas,0}$ ($p_{meas,1}$) for the $\ket{0}$ ($\ket{1}$) state.}
    \label{fig:readout_error_asymmetry_effect}
\end{figure}

In Figure~\ref{fig:Asymmetry-ratio_distr} we demonstrate how on the \textit{ibmq\_manhattan} device the two readout errors $p^r_0, p^r_1$ are correlated. 
The majority of readout error values correspond to $p^r_0 \lesssim 0.02$. 
For these points the asymmetry ratio  $\frac{p^r_1}{p^r_0}$ is between 2 and 3. 
In Figure~\ref{fig:readout_error_asymmetry_effect} the difference in logical error computed with and without considering readout asymmetry is presented for a $4+1$-qubit split encoding, assuming a readout asymmetry of 3. 
In case of the $\ket{0}$-state encoding the error due to neglecting the readout-asymmetry is increasing linearly in the range $p^r_0 \in [0,0.02]$ from $0$ to less than $0.002$.
Similarly, the effect of neglecting the readout-asymmetry when encoding the $\ket{1}$-state ranges from $0$ to less than $0.01$ for $p_1^r \in [0,0.2]$. 
For typical readout error values, such as $p_0^r=0.01$ and $p_1^r=0.03$, this leads to an absolute error due to the approximation of less than $0.001$ for the $4+1$-qubit split encoding, justifying the approximation of neglecting the asymmetry. 
We used the split encoding as a demonstrative example, but the same behavior is expected also for the other repetition encodings.

\section{CNOT error distribution}
\label{app:cnot_error_distribution}

\begin{figure}[ht!]
    \centering
    \includegraphics[width=\linewidth]{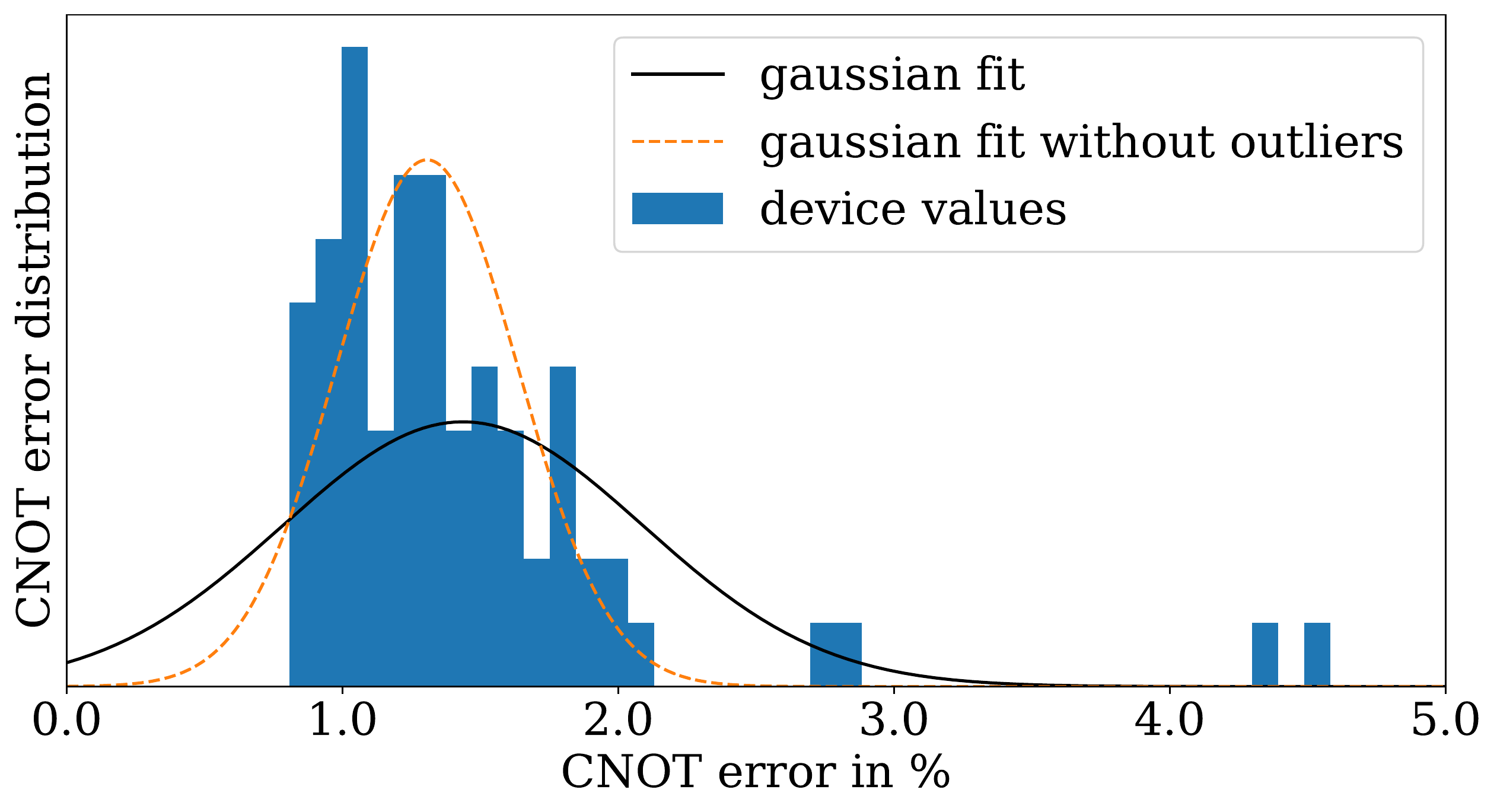}
    \caption{Histogram of the CNOT-errors of the $\textit{ibmq\_manhattan}$ device. Two Gaussian fits are shown, in black the one for all CNOT-errors, for the fit in orange the two outliers at $\sim4.5\%$ are ignored.}
    \label{fig:Cnot_err_distr_manhattan}
\end{figure}

The hypothesis of a gaussian distribution for CNOT errors does not fully agree with the hardware behavior, but still can be considered a good approximation for modelling the noise. 
In Figure \ref{fig:Cnot_err_distr_manhattan} we highlight the corresponding distribution for \textit{ibmq\_manhattan}. 
If one fits this distribution without considering the few outliers the fit seems to be more consistent and can describe more accurately the actual distribution of errors for the CNOTs. 

\section{Bloch sphere scan}

States on the Bloch sphere have been prepared and measured, both for a single physical qubit and a $10+1+1$ circular repetition encoded qubit. 
The qubits are the same as used for the $H_2$ calculation, i.e. one hexagon on the \textit{ibmq\_manhattan} device, with qubit $33$ on the device being both the single physical qubit and the root qubit for the encoding, and qubit $47$ serving as the flag qubit in the circular encoding (see Fig.~\ref{fig:all_hardware}c). 
Around each axis of the Bloch sphere $21$ equidistant rotations with angles $\theta_X, \theta_Y, \theta_Z$ from $0$ to $\pi$ have been performed. 
Then, the resulting states were measured in the computational basis. 
The initial state is $\ket{0}$, however in case of the $R_Z$ rotations a $R_X(\frac{\pi}{2})$ rotation precedes the $R_Z$ rotation in order to move along the equator of the Bloch sphere. 
Each of the $3\cdot21=63$ circuits that prepare and measure the points on the Bloch sphere were prepared and measured $57344$ times.

\begin{figure*}[]
    \centering
    \includegraphics[width=0.95\textwidth]{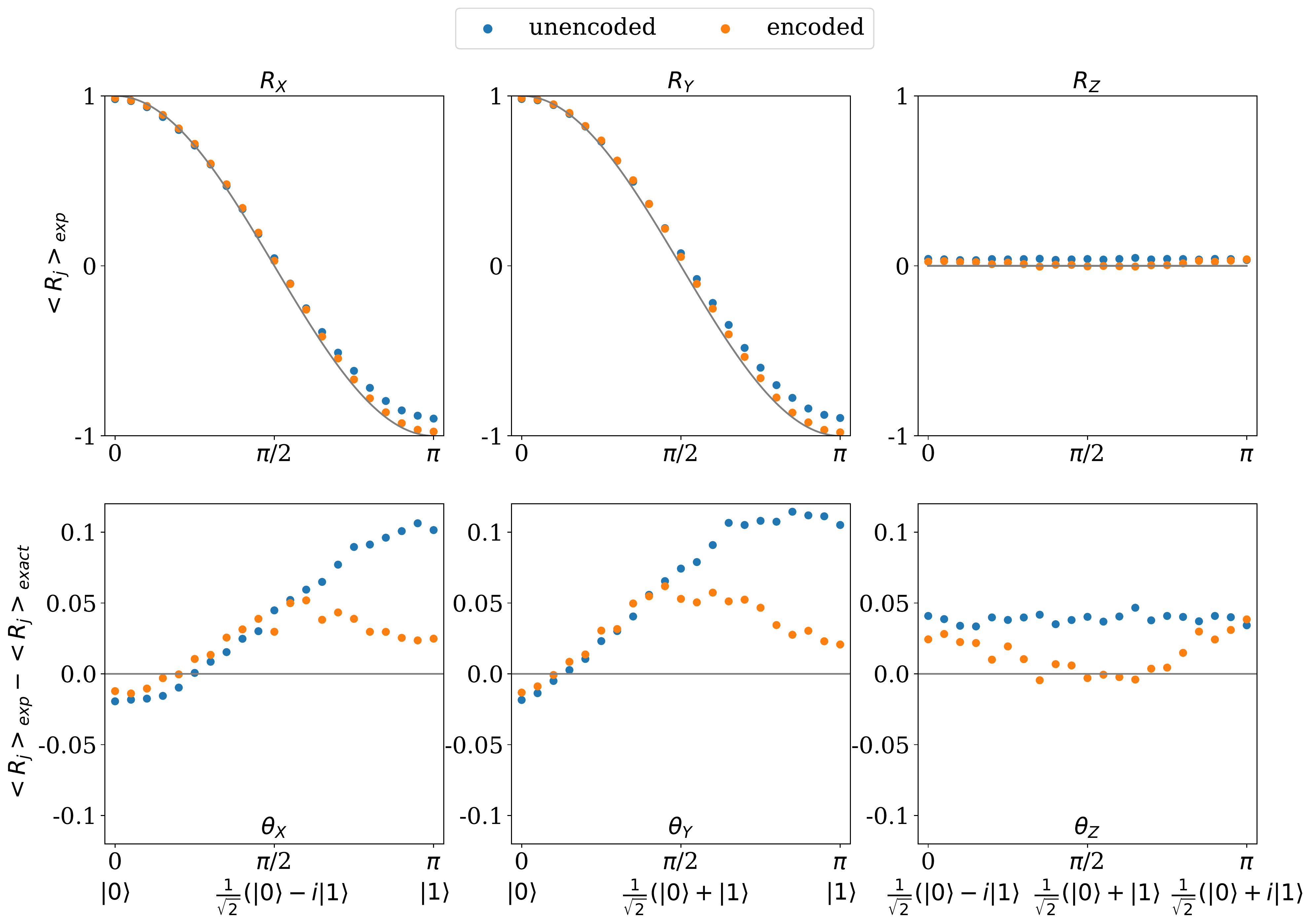}
    \caption{Expectation values of unencoded (blue) and circular encoded (orange) single-qubit rotations $R_X, R_Y, R_Z$ are shown for rotation angles in the interval $[0, \pi]$. The solid black line shows the exact value. The plots in the bottom row show the difference between the experimental data points and the exact values.}
    \label{fig:Bloch_sphere_scan}
\end{figure*}

As depicted in grey in Fig.~\ref{fig:Bloch_sphere_scan}, the expected outcomes describe cosines for the $R_X$ and $R_Y$ rotations and a constant $0$-function for the $R_Z$ rotation. 
Assuming the errors during measurement dominate other error sources for such short single qubit circuits, the lower row offers a good description of the readout errors for various points on the Bloch sphere. 
In line with the results from Section~\ref{sec:benchmarking_rep_codes} of the main text, the plots of the $R_X$ and $R_Y$ rotations illustrate how the circular encoding reduces readout errors the better the closer $\theta_X$ and $\theta_Y$ are to $\pi$, i.e. the higher the amplitude of $\ket{1}$ is for the prepared state. 
It is worth noting that the readout error of the encoded qubit for equal superpositions of $\ket{0}$ and $\ket{1}$ is not constant, but performs a small oscillation between expectation values of $0$ and $0.05$. 
Furthermore it seems as if for such superposition states the readout error of the encoded qubit is not independent on how the point on the Bloch sphere was reached. 
As an example the rotations $\theta_Y=\frac{\pi}{2}$ and $\theta_Z=\frac{\pi}{2}$ both should prepare the state $\frac{1}{\sqrt{2}}(\ket{0}+\ket{1})$, however the $R_Y(\frac{\pi}{2})$ rotation leads to a readout error of about $0.05$ compared $0$ for the $R_Z(\frac{\pi}{2})$ rotation. 
Reasons for this were not investigated in this study, but the discrepancy arises possibly from the hardware-implementation of the circuit and the different sources of noise that are beyond our consideration. 

\section{IBM Quantum hardware}
\label{app:ibmq_hardware}

\begin{figure}[ht!]
    \centering
    \includegraphics[width=0.9\linewidth]{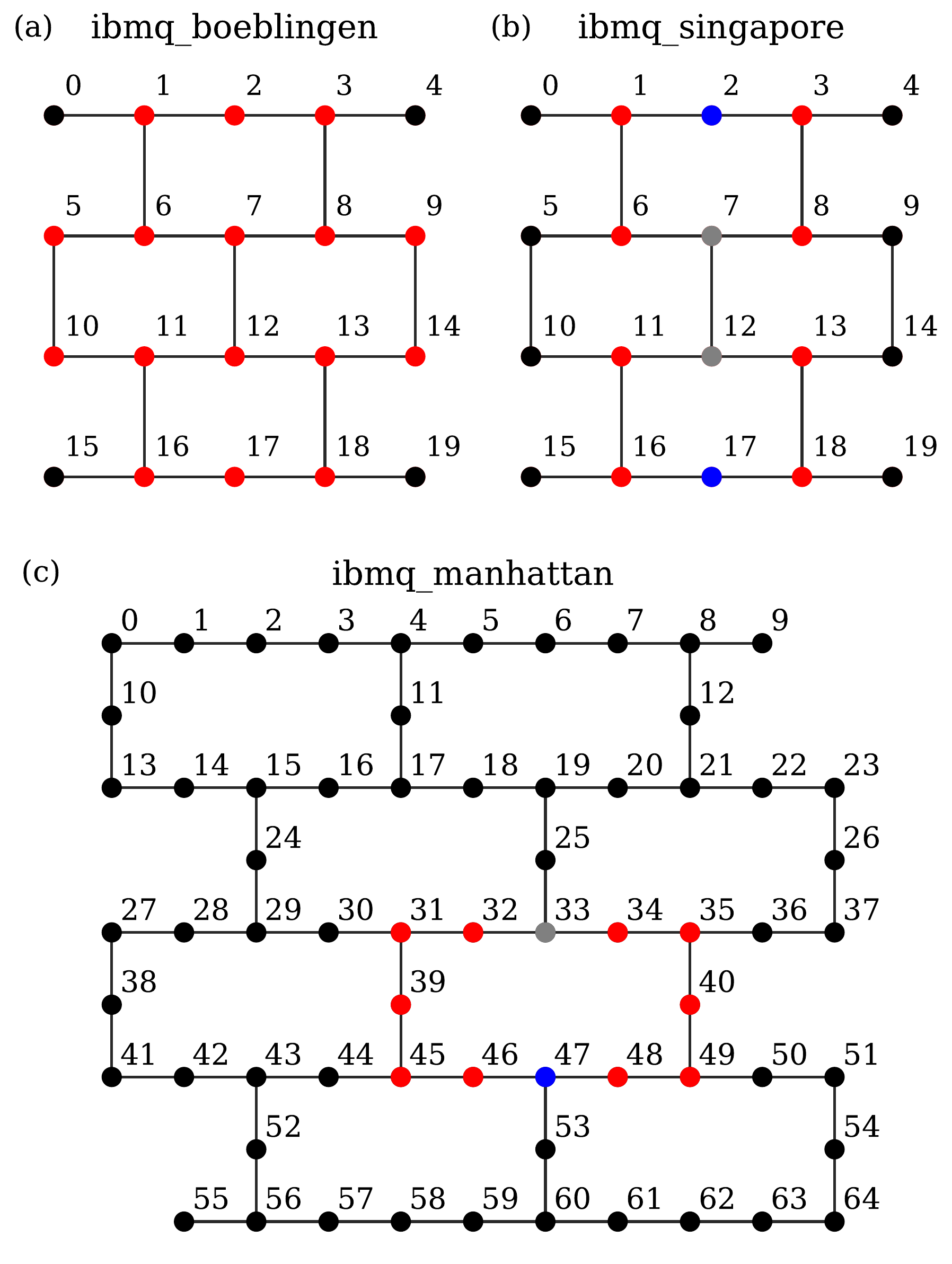}
    \caption{Connectivity for the different IBM Quantum processors (a) \textit{ibmq\_boeblingen}, (b) \textit{ibmq\_singapore} and (c) \textit{ibmq\_manhattan} quantum processors.The colored qubits are used as physical, repetition or flag qubits to produce the results of this work. For \textit{ibmq\_manhattan} we highlight the configuration of qubits used for the calculation of $\rm{H_2}$. For the benchmarking of Fig.~\ref{fig:benchmark_all_encodings_hardware}b of the main text all 12 possible rotations of the layout within the same heavy hexagon were used.}
    \label{fig:all_hardware}
\end{figure}

For our results we use many quantum processors provided via cloud from IBM Quantum.
Information about all the processor are provided via Qiskit~\cite{Qiskit_short}. 
For Figure~\ref{fig:benchmark_all_encodings_hardware} of the main text, we tested all 24 possible 6-qubit circular encodings for the \textit{ibmq\_boeblingen} device, whereas for the \textit{ibmq\_manhattan} we keep the same 12 qubits (31,32,33,34,35,39,40,45,46,47,48,49) and use each of them as a root qubit. 
The same qubits from \textit{ibmq\_manhattan} were used for the simulation of H$_{\rm{2}}$ molecule. 
The dynamics of the Ising model were executed in \textit{ibmq\_singapore} using qubits 7 amd 12 as root qubits and following the circular encoding as shown in Figure~\ref{fig:all_hardware}b.

\end{document}